\documentstyle[aasms4]{article}

\input psfig
\def\puncspace{\ifmmode\,\else{\ifcat.\C{\if.\C\else\if,\C\else\if?\C\else%
\if:\C\else\if;\C\else\if-\C\else\if)\C\else\if/\C\else\if]\C\else\if'\C%
\else\space\fi\fi\fi\fi\fi\fi\fi\fi\fi\fi}%
\else\if\empty\C\else\if\space\C\else\space\fi\fi\fi}\fi}
\def\SP{\let\\=\empty\futurelet\C\puncspace }
\def\etal{et\SP al.\SP }

\def\h-1{$h^{-1}$}

\def\void#1{{}}

\def\h1{$h^{-1}$}

\def\kms{kms$^{-1}$ }
\def\etal{et al.\,}
\def\eg{e.g., \,}

\def\lsim{~\rlap{$<$}{\lower 1.0ex\hbox{$\sim$}}}
\def\gsim{~\rlap{$>$}{\lower 1.0ex\hbox{$\sim$}}}
\def\hmu2 {$\bar{\mu}_{1/2}$\SP}
\def\mue {$\bar{\mu}_e$\SP}

\def\bri {mag~arcsec$^{-2}$\SP}

\def\m26 {$m_{26}$\SP}
\def\d26 {$D_{26}$\SP}

\def\mssrs2 {$m_{ssrs2}$\SP}

\def\AJ#1,{AJ, {#1},}
\def\ApJ#1,{ApJ, {#1},}
\def\ApJL#1,{ApJ, {#1},}
\def\ApJS#1,{ApJS, {#1},}
\def\MN#1,{MNRAS, {#1},}
\def\MNRAS#1,{MNRAS, {#1},}
\def\PASP#1,{PASP, {#1},}
\def\Science#1,{Science, {#1},}
\def\Nature#1,{Nature, {#1},}
\def\ZfA#1,{{\sl Zeitschrift f\"ur Astrophysik}, {#1},}

\def\dnsig{$D_n-\sigma$\SP}

\def\mg{Mg$_2$\SP}

\begin{document}

\title{Redshift-Distance Survey of Early-type Galaxies. I. Sample
Selection, Properties and Completeness \footnote{Based on observations
at Complejo Astronomico El Leoncito (CASLEO), operated under agreement
between the Consejo Nacional de Investigaciones Cient\'\i ficas de la
Rep\'ublica Argentina and the National Universities of La Plata,
C\'ordoba and San Juan; Cerro Totolo Interamerican Observatory (CTIO),
operated by the National Optical Astronomical Observatories, under
AURA; European Southern Observatory (ESO), partially under the ESO-ON
agreement; Fred Lawrence Whipple Observatory (FLWO); Observat\'orio do
Pico dos Dias, operated by the Laborat\'orio Nacional de Astrof\'\i
sica (LNA); and the MDM Observatory at Kitt Peak} }

\author{L. N. da Costa \altaffilmark{1,}\altaffilmark{2},
M. Bernardi\altaffilmark{1,}\altaffilmark{3,}\altaffilmark{4},
M. V. Alonso\altaffilmark{5}, G. Wegner\altaffilmark{6}}

\author{C. N. A. Willmer\altaffilmark{2,}\altaffilmark{7},
P. S. Pellegrini\altaffilmark{2}, C. Rit\'e\altaffilmark{2},
M. A. G. Maia\altaffilmark{2}}

\affil{\altaffiltext{1}{European Southern Observatory,
Karl-Schwarzschild Strasse 2, D-85748 Garching,
Germany}}

\affil{ \altaffiltext{2} {Departamento de Astronomia,
Observat\'orio Nacional, Rua General Jos\'e Cristino
77, Rio de Janeiro, R. J., 20921, Brazil}}

\affil{\altaffiltext{3}{Universit\"{a}ts-Sternwarte M\"{u}nchen,
Scheinerstr. 1, D-81679, M\"{u}nchen, Germany}}

\affil{\altaffiltext{4}{Max Planck Institut f\"ur Astrophysik, 
Karl-Schwarzschild Strasse 1, D-85740, Garching, Germany}}

\affil{\altaffiltext{5}{Observatorio Astr\'onomico de
C\'ordoba,  Laprida  854, C\'ordoba, 5000, Argentina}}

\affil{\altaffiltext{6}{Department of Physics \& Astronomy, Dartmouth
College, Hanover, NH  03755-3528, USA}}

\affil{ \altaffiltext{7}{UCO/Lick Observatory, University of California,
1156 High Street, Santa Cruz,  CA 95064, USA}}


\begin{abstract}

This is the first in a series of papers describing the recently
completed all-sky redshift-distance survey of nearby early-type
galaxies (ENEAR) carried out for peculiar velocity analysis. The
sample is divided into two parts and consists of 1607 elliptical and
lenticular galaxies with $cz\leq7000$ \kms and with blue magnitudes
brighter than $m_B=14.5$ (ENEARm), and of galaxies in clusters
(ENEARc).  Galaxy distances based on the \dnsig and Fundamental Plane
(FP) relations are now available for 1359 and 1107 ENEARm galaxies,
respectively, with roughly 80\% based on new data gathered by our
group. The \dnsig and FP template distance relations are derived by
combining 569 and 431 galaxies in 28 clusters, respectively, of which
about 60\% are based on our new measurements.

To date the ENEAR survey has accumulated: 2200 $R-$band images
yielding photometric parameters for 1398 galaxies; and 2300 spectra
yielding 1745 measurements of central velocity dispersions and
spectral line indices for 1210 galaxies.  In addition, there are some
1834 spectra of early-type galaxies available in the SSRS+SSRS2
database out of which roughly 800 galaxies yield high-quality
measurements of velocity dispersions and spectral line indices,
bringing the total number of galaxies with available spectral
information to about 2000. Combined with measurements publicly
available, a catalog has been assembled comprising: $\sim$ 4500
measurements of central velocity dispersions for about 2800 galaxies;
$\sim$ 3700 measurements of photometric parameters for about 2000
galaxies; distances for about 1900 galaxies. This extensive database
provides information on galaxies with multiple observations from
different telescope/instrument configurations and from different
authors. These overlapping data are used to derive relations to
transform all available measurements into a common system, thereby
ensuring the homogeneity of the database.

The ENEARm redshift-distance survey extends the earlier work of the 7S
and the recent Tully-Fisher surveys sampling a comparable volume. In
subsequent papers of this series we intend to use the ENEAR sample by
itself or in combination with the SFI Tully-Fisher survey to analyze
the properties of the local peculiar velocity field and to test how
sensitive the results are to different sampling and to the distance
indicators. We also anticipate that the homogeneous database assembled
will be used for a variety of other applications and serve as a
benchmark for similar studies at high-redshift.

\end {abstract} 

\section{Introduction}

Measurements of the peculiar motions of galaxies in the nearby
universe represent one of the most powerful tools available to probe
mass fluctuations on scales $\lsim 100 {\rm h^{-1}}$~Mpc,
complementing current efforts to measure the amplitude of these
fluctuations from one-degree scale observations of the cosmic
microwave background and on very large scales as obtained from the
analysis of COBE data (\eg G\'orski \etal 1996).  Analysis of the
peculiar velocity field provides not only a direct test of the
gravitational instability picture but it also enables the relationship
between galaxies and the underlying mass distribution to be
investigated. Moreover, to the extent that this relationship can be
adequately modeled, for instance by a linear biasing model, comparison
between peculiar velocity and redshift data can be used to determine
the parameter $\beta=\Omega^{0.6}/b$, where $\Omega$ and $b$ are the
density and linear biasing parameters, respectively (\eg Dekel 1994;
Strauss \& Willick 1995).

While the importance of cosmic flows to cosmological studies has long
been recognized, and a variety of methods for analyzing peculiar
velocity data have been developed over the years, the progress of
redshift-distance surveys suitable for such studies has been
relatively slow. Most of the work in the field relied until recently
in the Tully-Fisher (TF) and $D_n-\sigma$ samples of Aaronson \etal
(1982) and Lynden-Bell \etal (1988, hereafter 7S), respectively,
complemented by relatively small samples covering particular regions
of the sky (Willick 1991; Courteau \etal 1993). Early analyses led to
important results such as the discovery of the Great Attractor by the
7S and to the rejection of the high-bias CDM model as a viable
cosmological model. However, the sparseness of the samples and the
relatively small effective volume they probe also led to results and
interpretations that need further confirmation. These include large
derived values of the cosmological density parameter, at variance with
other estimates, and the existence of large scale coherent motions,
suggestive of excess power at very large scales (Willick 1990;
Mathewson, Ford \& Buchhorn 1992, hereafter MFB; Courteau \etal
1993).

Recently, the observational situation has significantly improved with
the completion of large TF surveys in both hemispheres (MFB; Mathewson
\& Ford 1996; Giovanelli \etal 1997~a,b; Haynes \etal
1999~a,b). Combined, they provide the largest all-sky sample currently
available for peculiar velocity studies greatly extending the depth
and the number of galaxies of earlier surveys. There still are,
however, points of concern.

First, these surveys were carried out independently with different
selection criteria, different observing and reduction techniques and
have been assembled in different ways to produce all-sky catalogs
suitable for analysis such as the SFI (\eg da Costa \etal 1996;
Giovanelli \etal 1998) and Mark~III (Willick \etal 1997)
catalogs. These catalogs differ both in the data sets used to assemble
them and in the transformations employed to map the MFB data in the
Southern hemisphere onto a common system.  Analysis of these catalogs
have led to both consistent (\eg Zaroubi \etal 1996; Freudling \etal
1999) and conflicting results (\eg Nusser, Davis \& Willick 1997; da
Costa \etal 1998a; Borgani \etal 1999; Willick \& Strauss 1998)
depending on the type of analysis considered.

Second, most redshift-distance surveys of the whole sky currently
available, including the recently completed Shellflow survey (Courteau
\etal 1999), rely primarily on TF distances.  While other recent
investigations that have used early-type galaxies, have either focused
on motions of clusters of galaxies (\eg EFAR, Wegner \etal 1996,
Colless \etal 1999; SMAC, Hudson \etal 1999; LP10K, Willick
1999) or have concentrated on pre-chosen areas of the sky which do not
render them optimal for all types of analyses (\eg M\"uller \etal
1998; M\"uller, Wegner \& Freudling 1999). Finally, recent attempts to
use more accurate distance indicators, such as the SBF method (Tonry,
Blakeslee, \& Dressler 1999) and SNeIa (\eg Riess 1999),
remain limited in depth and in the number of objects with measured
distances to be used for a detailed mapping of the peculiar velocity
field (but see Blakeslee \etal 1999).

In order to address some of these concerns we have carried out the
so-called ENEAR survey, an extensive spectroscopic and $R-$band
imaging survey of a sample of nearby early-type galaxies (thus the
choice of the name to contrast with the EFAR project) brighter than
$m_B=14.5$ and $cz\leq7000$\kms (hereafter ENEARm), extracted from
complete redshift surveys, and of galaxies in 28 selected clusters
(ENEARc). Currently, the ENEARm and ENEARc samples comprise 1359 and
569 galaxies with measured distances, respectively. More importantly,
82\% of the ENEARm galaxies and 58\% of the ENEARc galaxies have
distance estimates based on new measurements obtained by our group in
both hemispheres, making it the largest homogeneous all-sky survey of
its kind conducted by a single group.  The primary goal of this survey
has been to extend the 7S sample to fainter magnitudes, thereby
probing a volume comparable to that of existing TF surveys such as the
SFI. With such a sample it is possible to: 1) investigate the
reproducibility of the results obtained from the analysis of peculiar
velocity data as measured by different secondary distance indicators;
2) combine it with the TF surveys to produce a homogeneous all-sky
catalog of galaxies of different morphological types, sampling both
low and high density regions; and 3) provide a denser sampling of the
gravitational field.

The ENEAR survey began as an outgrowth of spectroscopic surveys
carried out in the Southern hemisphere started in 1982 (\eg
SSRS+SSRS2, da Costa \etal 1988, 1998b). Although these surveys were
primarily designed for redshift measurements, from the start an
attempt was made to obtain higher signal-to-noise spectra for bright
early-type galaxies for their eventual use in the measurement of
central velocity dispersions and spectral line indices. The original
goal was to provide spectral information for a magnitude-limited
sample which could be used as a basis for the statistical analysis of
the properties of this population as a function of the local galaxy
density and different types of environment. By 1988, motivated by the
results of the 7S analysis and the start of the SFI TF-survey
attention shifted to the peculiar velocity field and $R$-band imaging
was initiated for a sample selected from a combination of
magnitude-limited samples with complete redshift information available
at the time. This sample was later extended to lower galactic
latitudes as described below. In the meantime, several galaxies with
only older spectra were re-observed to ensure the quality and
uniformity of the sample. In 1993 the campaign to observe northern
galaxies began mostly at the MDM Observatory.

As a result we have accumulated a large number of spectra and images
from which we have measured redshifts, central velocity dispersions,
spectral line indices and several photometric parameters.  We have
assembled this information together with spectroscopic and photometric
parameters available in the literature, to produce a homogeneous
catalog of early-type galaxies, hereafter ENEAR catalog. New data
continue to be recorded in the catalog as the observations of
early-type galaxies are still in progress. We hope to make our
database accessible on-line in the future.

The large number of galaxies in the ENEAR catalog, most of which with
new high-quality spectra and CCD images, its homogeneity and the
improved membership assignment of galaxies to groups/clusters for a
significant fraction of the galaxies make this catalog a valuable tool
for many other applications. The ENEAR catalog, currently the largest
uniform database of its kind, is ideally suited for: the statistical
characterization of the properties of present-day early-type galaxies;
stellar populations studies using different spectral line indices;
testing the universality of distance relations such as the \dnsig and
the Fundamental Plane (FP); assessing the influence of different types
of environment and the dependence of the properties of these galaxies on
the local density of galaxies; and constraining models for the formation
and evolution of ellipticals (\eg Bernardi \etal 1998). We also
anticipate that the present sample will be used as a benchmark to
comparable high-redshift studies.

The ENEAR survey represents a significant improvement over previous
redshift-distance surveys of early-type galaxies. Besides the larger
number of galaxies, about a factor of three larger than the 7S sample,
most of the galaxies have new measurements of velocity dispersion, in
general obtained from higher resolution spectra, and photometric
parameters derived from the analysis of $R$-band images. To ensure
uniformity, a large number of observations have been obtained of
galaxies with measurements available from other authors or from
different telescope/instrument configurations enabling all available
measurements to be scaled to a uniform system. Since the survey has
been conducted by a single group, all of the data were analyzed using
the same procedure, further ensuring their uniformity, and the
observations were coordinated so as to reach the same level of
completeness over the whole sky. Finally, the homogeneous and
well-defined sample suitable for peculiar velocity analysis is a
subset of magnitude-limited samples for which complete redshift
information is available from surveys such as CfA1+CfA2 (Huchra \etal
1983; Geller \& Huchra 1989; Falco \etal 1999), SSRS+SSRS2 and ORS
(Santiago \etal 1995), thus allowing an objective assignment of
galaxies to groups/clusters.

This paper is the first in a series reporting the results of the ENEAR
survey. Its primary goal is to give an overview of the project and
provide the basic information which will be used in future papers of
this series. In Section~\ref{selection}, we provide details of the
procedure adopted in the selection of a well-defined sample for
peculiar velocity analysis and of its general characteristics. We also
discuss the selection of the cluster sample, which will be used in
subsequent papers to construct template distance relations. In
Section~\ref{data}, we briefly describe our new spectroscopic and
$R$-band imaging data. We also discuss the procedure adopted for the
homogenization of the data either newly obtained with different
telescopes/instrument configurations or available in the
literature. Such a procedure is of paramount importance in the
elimination of inconsistencies between data from different sources in
order to minimize systematics. In addition, we summarize the data
currently available in the ENEAR catalog that has been assembled in
the course of this project. In Section~\ref{redshiftdistance}, we
briefly discuss the method employed to estimate galaxy distances, the
procedure adopted for pruning the sample of galaxies unsuitable for
peculiar velocity studies and describe the criteria employed for
assigning galaxies to groups/clusters. This grouping is crucial for
reducing the influence of virial motions and decreasing
distance-errors. We also discuss the completeness of our
redshift-distance survey, compare our sample with those of previous
peculiar velocity studies and give a preview of the measured peculiar
velocity field. Finally, a brief summary is presented in
Section~\ref{summary}.

\section{Sample Selection}
\label{selection}

\subsection{ENEARm:  The magnitude-limited sample}
\label{enearm}

As pointed out above the primary goal of the ENEAR project has been to
extend the volume probed by the 7S sample, which considered early-type
galaxies brighter than $m_B\sim13.5$ sampling the peculiar velocity
field within an effective volume $\sim$ 4000~\kms in radius. An
equally important results is that this complements the SFI TF
redshift-distance survey of late spirals (\eg Haynes \etal 1999a,b)
out to $\lsim$ 7000~\kms. 

To achieve our goal we assembled all the available complete redshift
surveys at the beginning of this project (in 1988) to construct an
all-sky sample. Originally the following samples were used: 1) the
CfA1 sample (Huchra \etal 1983), covering the regions $b>40^\circ$ and
$\delta >0^\circ$ in the northern galactic cap, and $b<-30^\circ$ and
$\delta >-2.5^\circ$ in the southern hemisphere; 2) the sample of
galaxies used in the SSRS (da Costa \etal 1988) covering the regions
$b<-30^\circ$ and $\delta\leq-17.5^\circ$ in the southern galactic
cap, and $b>40^\circ$ and $\delta <-17.5^\circ$ in the northern
galactic cap; 3) the equatorial survey sample of Huchra \etal (1993)
filling in the gaps between the first two samples near the
equator. The samples for these redshift surveys were extracted from
different catalogs which include: The Surface Photometry Catalogue of
the ESO-Uppsala Galaxies (ESO, Lauberts \& Valentjin 1989), the
Morphological Catalog of Galaxies (MCG, Vorontsov-Velyaminov \&
Arhipova 196), the Uppsala General Catalog (UGC, Nilson 1973) and the
Catalog of Galaxies and Clusters of Galaxies (CGCG, Zwicky \etal
1961-1968). Magnitudes for the SSRS galaxies were assigned based on
magnitude-diameter relations derived by Pellegrini \etal (1990) who
made the first attempt to construct a uniform magnitude-limited sample
for the whole sky, with additional checks on the relative differences
in the magnitudes of the different catalogs being conducted later
(Alonso \etal 1993, 1994). In the southern hemisphere the ENEAR
magnitudes have been compared to the more recent SSRS2 magnitude
system (da Costa \etal 1998b). We find that the magnitudes obtained
from the magnitude-diameter relation show a linear relation with a
scatter of $\sim 0.45$~mag. More recently, galaxies drawn from the ORS
(Santiago \etal 1995) were added in order to extend the sample towards
lower galactic latitudes and cover at least part of the Great
Attractor region.

Magnitudes used in the different catalogs were converted into an
approximately homogeneous system using statistical corrections between
the various systems as described in Pellegrini \etal (1990) and da
Costa \etal (1998b). However, as discussed in these papers the
individual magnitudes can have errors as large as 0.5~mag.  The
morphological types adopted in our final catalog are those from
Lauberts \& Valentijn (1989), and the types given in the other
catalogs have been converted into this system.  From the experience
accumulated in the morphological classification of the SSRS2 (da Costa
\etal 1998b) we know that morphological misclassifications are
frequent and in the course of the observations several objects had to
be removed both before, based on the examination of the digitized sky
survey (DSS, Lasker \etal 1990), and after the observations from the
examination of the gathered CCD images, as discussed below.

\begin{figure}
\centering 
\mbox{\psfig{figure=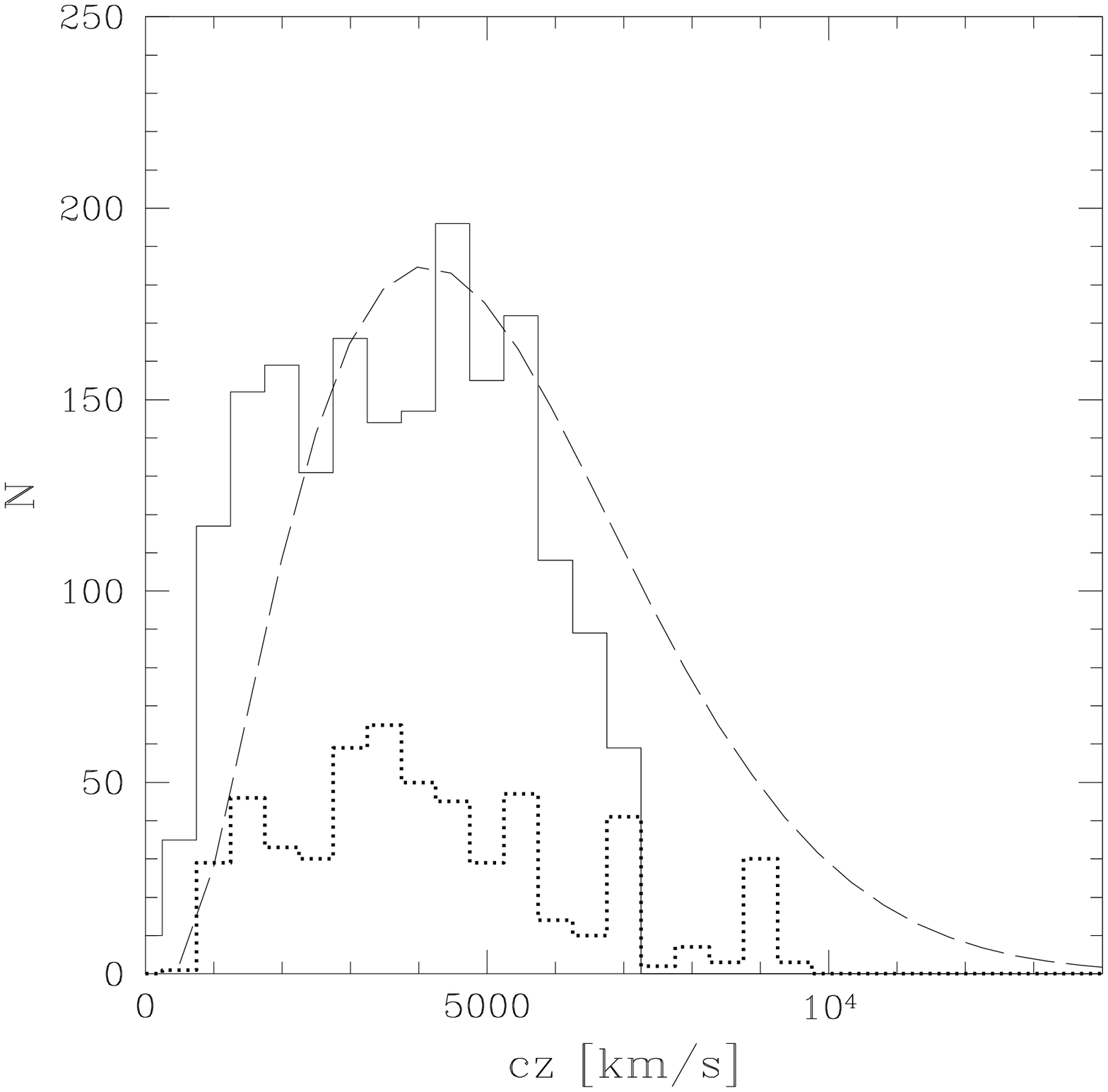,height=10truecm}}
\caption{Redshift distribution of galaxies in the magnitude and
redshift-limited ($cz\leq 7000$~\kms) ENEARm sample (solid histogram)
compared to that of the 7S sample (dotted histogram).  The dashed line
is the redshift distribution predicted from a uniform distribution of
early-type galaxies with a luminosity function given by that
determined by Marzke \etal (1998) for early-type galaxies in the
SSRS2}
\label{fig:cztot}
\end{figure}

From the all-sky sample we have drawn 1847 galaxies brighter than
$m_b=14.5$, with morphological types $T\leq-2$ and $cz\leq7000$~\kms.
These criteria define the original ENEARm sample considered for peculiar
velocity studies, but as shown below it includes a significant number of
galaxies which, for different reasons, are not suitable for observations
or for estimating distances. The sample comprises 471 galaxies with
$T\leq-5$, 305 galaxies with $T=-3$, and 1071 with $T=-2$, but these
numbers should be viewed only as a rough indication of the morphological
mix of the catalog, given the uncertainties in the classification. The
redshift distribution of the sample is shown in Figure~\ref{fig:cztot},
where it is compared to that expected for a $m_B=14.5$ magnitude-limited
sample and to the redshift distribution of galaxies in the 7S sample.  The
predicted distribution was obtained using the luminosity function
parameters derived by Marzke \etal (1998) for early-type galaxies,
normalized to the area of the sky covered by the ENEARm sample. The
redshift cutoff was adopted to ensure that the sample depth was
comparable to that of the SFI, and as a trade-off between reaching a
high level of completeness in the nearby volume and the available
observing time. The ENEARm sub-sample corresponds to $\sim85\%$ of the
total number of early-types in a $m_B=14.5$ magnitude-limited sample.

\begin{figure}
\centering
\mbox{\psfig{figure=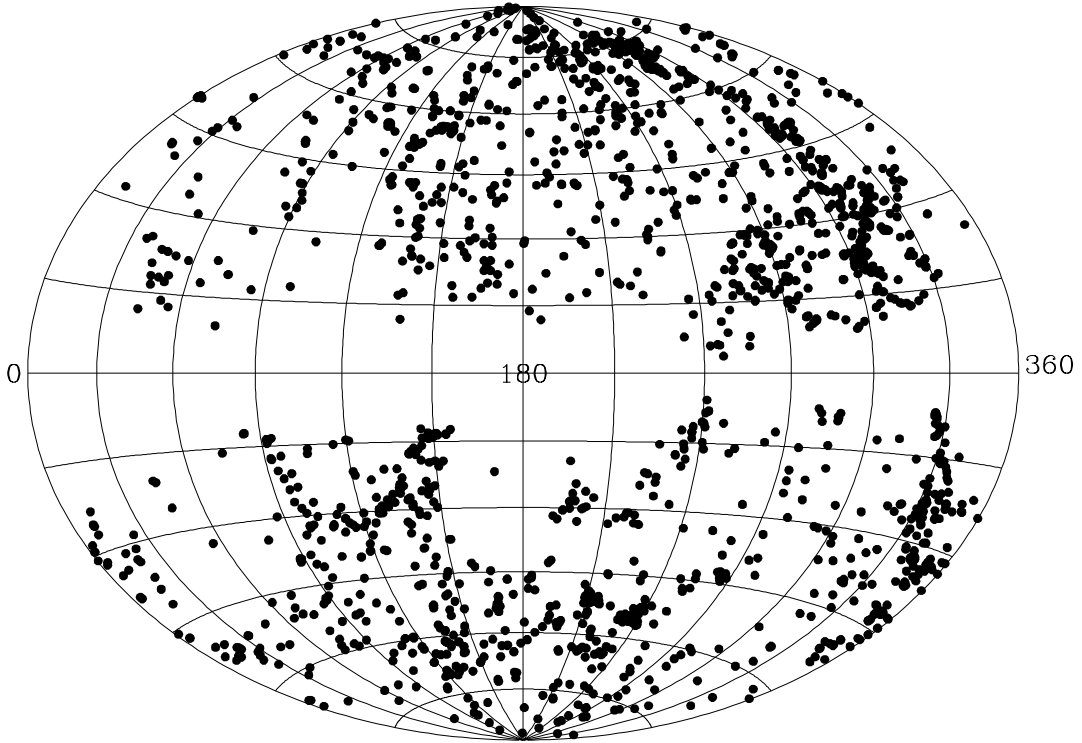,height=16truecm}}
\caption{The projected distribution in galactic coordinates of the
early-type galaxies in the magnitude and redshift-limited ENEARm sample
as extracted from the parent catalog before pruning galaxies unsuitable
for distance estimates (see text)}.
\label{fig:enear}
\end{figure}

The projected distribution of the ENEARm sample is shown in
Figure~\ref{fig:enear}. Comparison of this figure with that shown, for
instance, in Santiago \etal (1995) for an all-sky sample to about the
same depth shows that the ENEARm densely probes the most prominent
large-scale structures in the nearby universe including Virgo, the
Great Attractor region and its extension to the Telescopium-Pavo-Indus
supercluster, and the Perseus-Pisces complex.  Also note the presence
of areas devoid of galaxies which correspond to high-extinction
regions as shown by Santiago \etal (1995).

\subsection {ENEARc: The Cluster Sample}
\label{clusters}

The ENEARm sample is complemented by a cluster sample, hereafter
ENEARc, which has been used to define template distance relations
(Bernardi \etal 1999b) combining all available cluster data. The
sample consists of groups identified in ENEARm (see
section~\ref{group}) with more than 10 members and at least 5
early-type galaxies to which we have added clusters selected from the
literature. The latter were not identified as groups because either
most of the member galaxies are fainter than the ENEARm limiting
magnitude, have redshifts beyond its redshift limit, or are outside
the surveyed region at low galactic latitudes. In addition, to the
clusters identified as groups, fainter members have been added
according to a membership criteria similar to that described below
(see section~\ref{group}) and discussed in more detail by Bernardi
\etal (1999a).  A total of 37 groups were identified satisfying our
cluster definition, nearly all associated with Abell/ACO clusters.
The ENEARc sample currently comprises a total of 28 clusters, 22 of
these identified in the volume probed by ENEARm. For most of these
clusters a large number of modern observations are available in the
literature. This provides an adequate sample to derive distance
relations. Instead of observing new systems preference was given to
the observation of as many galaxies as possible already with available
data from different sources, primarily from J{\o}rgensen \etal (1992,
1995 a,b) and 7S.  This was done to enable us: to bring all of the
available data, both in clusters and in the field, to a common system;
to substitute old measurements of central velocity dispersions and of
photometric parameters, whenever required; to obtain better resolution
and higher $S/N$ spectra for galaxies in clusters for comparison with
similar data for galaxies in low density regions; to enlarge the
sample of galaxies with the parameters required for the derivation of
the FP; to enlarge the sample of galaxies in clusters with measured
distances to improve the statistics of the template relation.

\begin{figure}
\centering
\mbox{\psfig{figure=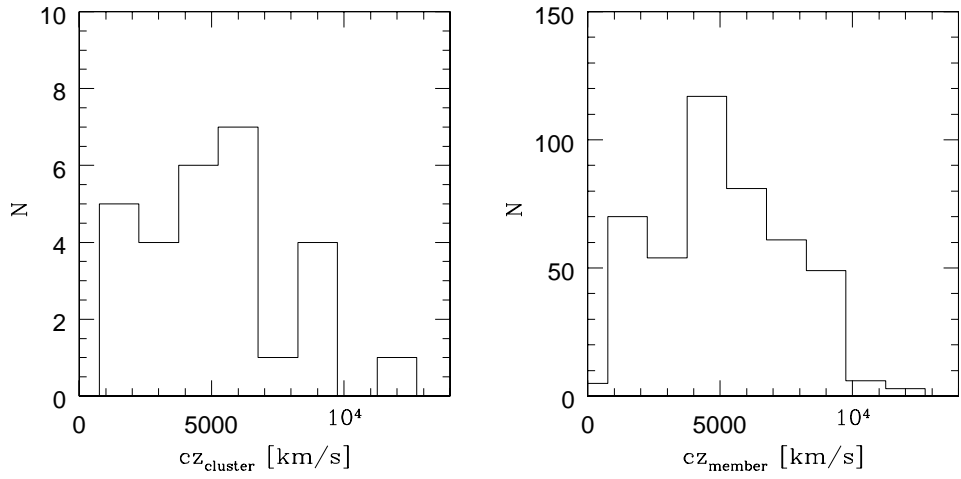,height=10truecm,bbllx=2truecm,bblly=13truecm,bburx=19truecm,bbury=27.5truecm}}
\caption{Redshift distribution of clusters (left panel) and cluster
members (right panel) in  the ENEARc sample.} 
\label{fig:czclust}
\end{figure}

\begin{figure}
\centering
\mbox{\psfig{figure=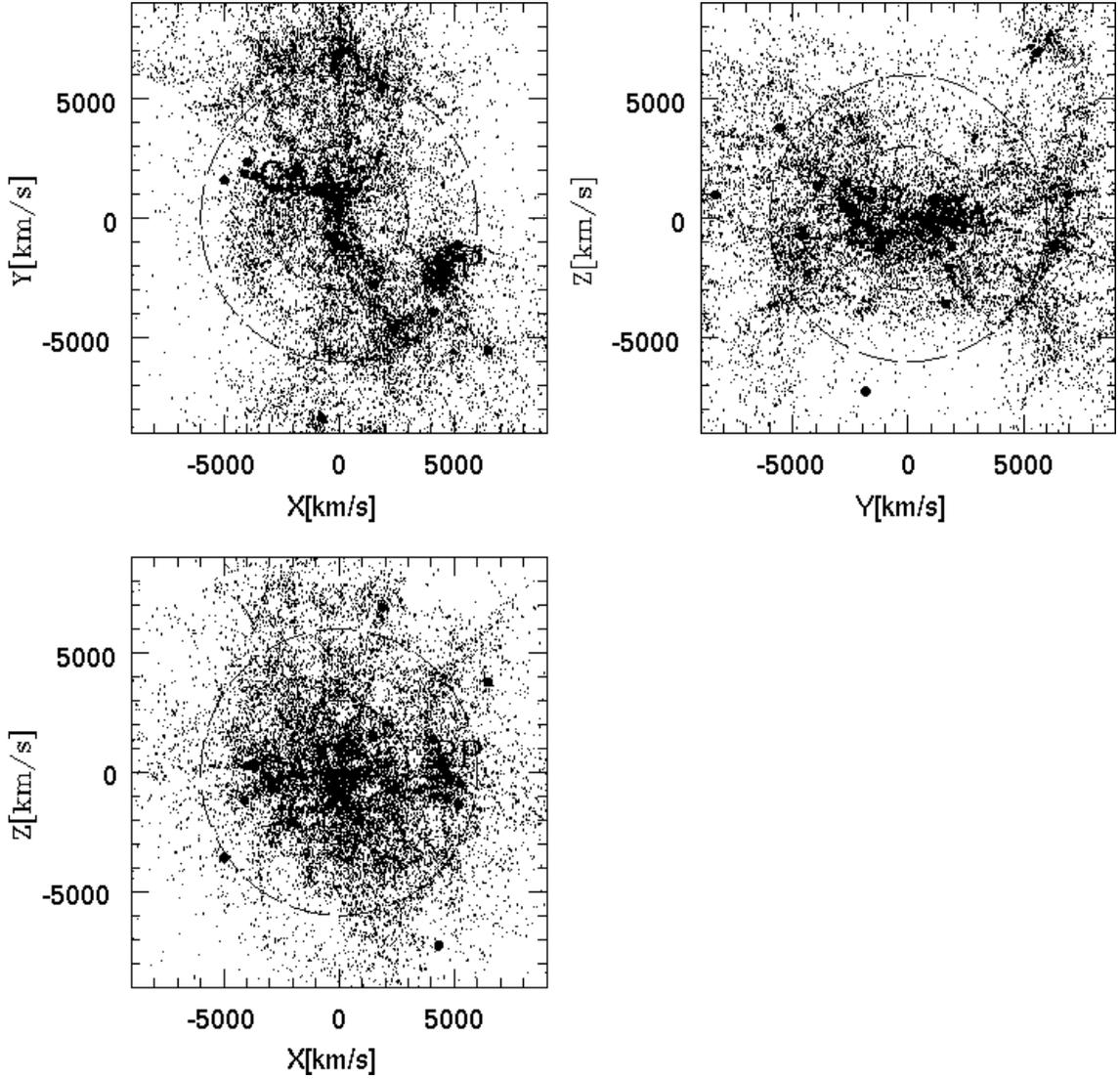,height=16truecm,bbllx=1.5truecm,bblly=5truecm,bburx=18.5truecm,bbury=25truecm
}} \caption{The spatial distribution of the 28 ENEARc clusters in
Cartesian Supergalactic coordinates (X, Y, Z), expressed in \kms in the
CMB  reference frame. The two dominant concentrations of galaxies,  the
Great Attractor (GA) and the Perseus-Pisces (PP) superclusters,  are
indicated on the three panels. Small dots show the objects  in the
parent magnitude-limited redshift survey samples from  which our catalog
is drawn.} 
\label{fig:sgcoor} 
\end{figure}

Figure~\ref{fig:czclust} shows the redshift distribution of the 28
clusters presented in Bernardi \etal (1999a) which span a range of
redshifts up to $cz \sim 10,000$~\kms with an approximately uniform
number per redshift bin. Also shown is the redshift distribution of
all the cluster members identified using the membership assignment
described below. The spatial distribution of these clusters is shown
in Figure~\ref{fig:sgcoor} where three orthogonal projections in
Supergalactic cartesian coordinates are displayed.

\section {Data}
\label{data}

As already stated, the primary objective of the present
redshift-distance survey has been to produce an all-sky homogeneous
sample for peculiar velocity analysis.  With a few exceptions, most
recent work in the field has had to rely on the combination of samples
from diverse sources, using different selection criteria, observing
strategies and reduction procedures. This has made the homogenization
of the measurements from different sources very difficult and their
errors poorly understood. By contrast, even though measurements from
different authors are used in compiling our final homogeneous
redshift-distance catalog, about 77\% and 65\% of early-type galaxies
with available photometric and spectroscopic data, respectively, are
from new observations obtained as part of this project over the whole
sky. This together with the large number of galaxies with multiple
observations from different telescope/instrument configurations and in
common with other authors enables the internal and external
homogenization of the whole data set, the measurement of parameters
with an improved accuracy and the possibility of cross-checks between
different data sets.

While the imaging observations described here are limited to the ENEAR
samples (ENEARm+ENEARc) presented above, spectra have been obtained
for galaxies belonging to the ENEARm and ENEARc samples as well as for
other early-type galaxies observed as part of several spectroscopic
surveys conducted in the Southern hemisphere (\eg SSRS, SSRS2).
Measurements of the central velocity dispersion and of spectral line
indices for galaxies with spectra deemed suitable for these
calculations have also been included in our ENEAR catalog presented
below to further enlarge the database of spectroscopic parameters
characterizing present-day early-type galaxies for studies other than
peculiar velocity analysis. In later papers in this series we will
present these additional measurements together with those obtained
from the ENEAR survey proper (Wegner \etal 2000; Rit\'e \etal 2000).

\subsection {ENEAR: Spectroscopic Observations}

Our spectroscopic observations have been carried out over a long
period of time using a variety of telescopes and instruments (CASLEO,
CTIO, ESO, LNA and MDM), detectors and gratings, with the spectral
resolution ranging from 2 to 5~\AA. Some of the data originating as
far back as the mid-80s were obtained using the intensified
photon-counting Reticon detector utilized in the Southern Sky Redshift
Survey (SSRS, da Costa \etal 1991). The observing and data reduction
procedures are similar to those used in the EFAR survey (Wegner \etal
1999).

To date a total of $\sim 2000$ spectra of early-type galaxies taken
from the ENEARm and ENEARc sample have been obtained. However, some of
these spectra have not yielded reliable measurements of the
spectroscopic parameters of interest (velocity dispersion; spectral
lines indices). These cases include strong emission-line galaxies with
weak absorption features, low surface brightness galaxies yielding low
$S/N$ spectra and galaxies too close to bright stars. Others,
especially the older Reticon data, lead to measurements with large
errors compared to our most recent observations and most have been or
are in the process of being re-observed.

Of the total number of available spectra, 1745 were suitable to
measure the central velocity dispersion of 1210 galaxies, including
535 multiple spectra of 327 galaxies.  The number of repeated
observations range from two to more than 10 for a few comparison
galaxies. The multiple observations were used to compare spectra
obtained with different telescope/instrument configurations and have
been used to make our measurements internally consistent. They have
also been used to calibrate our internal error estimates (Wegner \etal
2000). In addition, we have observed about 200 galaxies with
measurements available in the literature to derive statistical
corrections and bring those data into a uniform system (Bernardi \etal
1999a).

\begin{figure}
\centering
\mbox{\psfig{figure=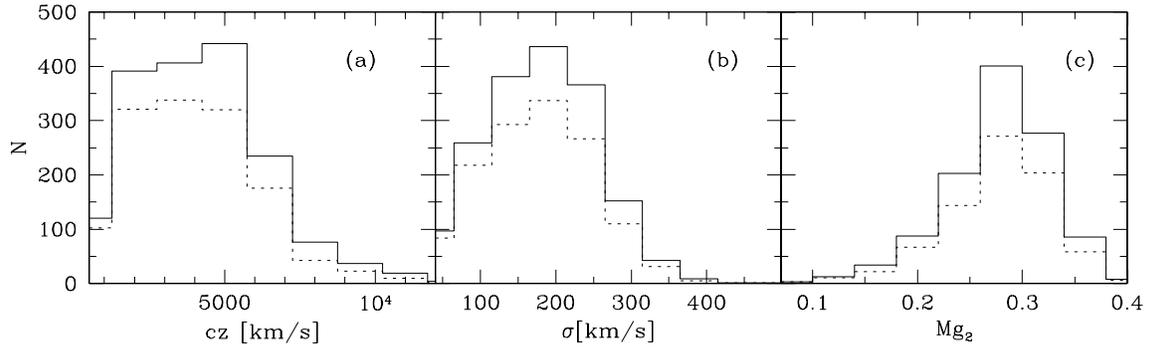,width=13truecm,bbllx=3truecm,bblly=18.5truecm,bburx=19truecm,bbury=25truecm }}
\caption{Distribution of redshift (panel a), central velocity
dispersion (panel b), and Mg$_2$ line index (panel c) for galaxies
observed by the ENEAR survey.  Dotted histograms represent
observations at high-resolution ($\lsim 2.5$\AA). The plots show the
total number of observations including multiple observations of the
same object.}
\label{fig:histspect}
\end{figure}

\begin{figure}
\centering
\mbox{\psfig{figure=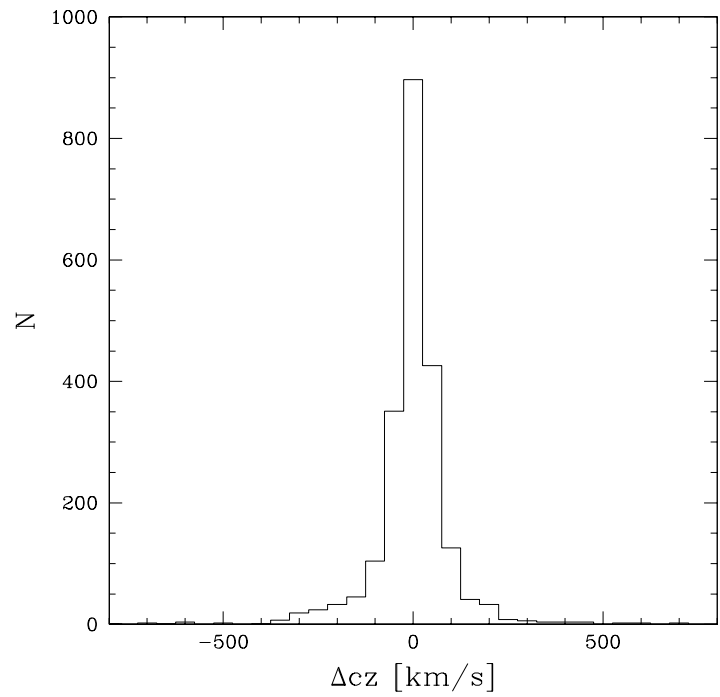,height=10truecm}}
\caption{Distribution of the differences between our new redshifts and
those previously available in the literature.}
\label{fig:histcompcz}
\end{figure}

By analyzing these spectra we have measured redshifts, velocity
dispersions and \mg line indices (Wegner \etal 2000) and
Figure~\ref{fig:histspect} shows the distribution of these quantities
for all ENEAR spectra deemed suitable for these measurements. Also
shown are the same distributions as measured from spectra obtained
using a resolution of $\lsim$2.5~\AA. Over 80\% of the spectra were
obtained at high-resolution allowing us to make more accurate
measurements, especially for galaxies with $\sigma \lsim 100$
kms$^{-1}$. The distribution of the differences between our redshifts
and those available in the literature is shown in
Figure~\ref{fig:histcompcz}. In general the agreement is excellent
leading to an rms of the differences $\sim$30~\kms. Note, however, the
presence of a few outliers which should be taken into account when
computing peculiar velocities. Since our measurements have been
obtained from high $S/N$ in computing peculiar velocities we give
preferences to these new measurements. Moreover, from our data we find
86 emission-line galaxies, out of which about 60 show no evidence of
being misclassified spirals. Some of these have very strong emission
lines and weak absorption features which may indicate the presence of
residual star-formation.  A detailed description of the observations,
data reduction, and derived spectroscopic parameters is given in
Wegner \etal (2000). Finally, we point out that in addition to the \mg
line index used by Bernardi \etal (1998), several other spectral
indices are being computed and will be presented in a future paper
(Rit\'e \etal 2000).

Our spectroscopic parameters have been converted onto a uniform system
using the many repeated observations we have in different runs in both
hemispheres. Our fiducial system is defined by our high-resolution
spectra ($\sim 2.5$~\AA). In addition, we used measurements of 200
galaxies in common with other sources to derive corrections to scale
the data in the public domain to our system.  From these external
comparisons we find that our errors are typically 8\% in $\sigma$ and
0.01~mag in \mg, consistent with our internal estimates (Wegner \etal
2000).

\subsection{ENEAR: Photometric Observations}

Since 1988, photometric observations in $R$-band have been made using
different telescopes in the northern (FLWO, MDM) and southern
hemispheres (CTIO, ESO).  About 2200 images have been obtained of
which 1896 were taken in photometric conditions. From these images we
have measured photometric parameters for 1398 galaxies.  Most of the
data were taken with large format CCDs, allowing for good sky
subtraction and, occasionally, for observing more than one galaxy per
frame, especially for galaxies in groups/clusters.  Some galaxies were
later discarded because they were unsuitable for measurements of the
photometric parameter of interest (\eg stellar contamination; crowded
fields; interacting galaxies or superposed galaxy images;
misclassification).

Because the imaging observations were conducted over an extended
period of time, an attempt was made to obtain repeated observations in
every run to provide the necessary data for run-to-run
corrections. Such observations have been used to transform all of our
measurements into a common internal system. 

\begin{figure}
\centering
\mbox{\psfig{figure=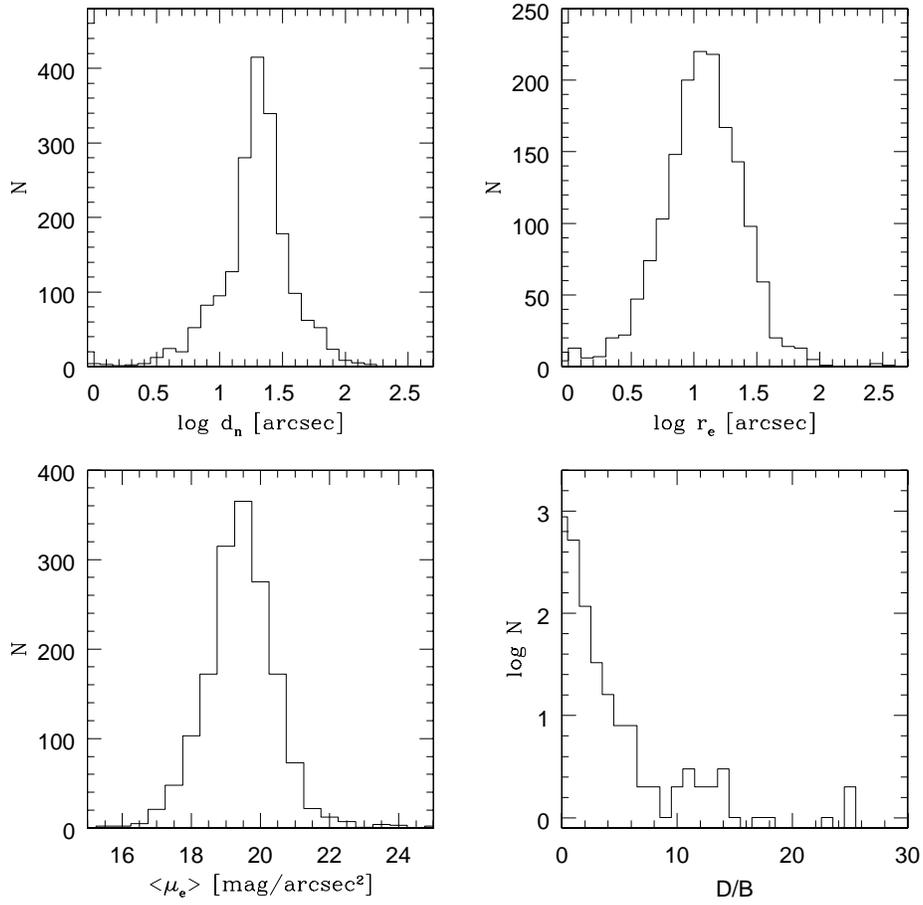,height=13truecm}}
\caption[Distribution of observed photometric parameters: 
$\log{d_n}$, $\log{r_e}$, 
$\bar{\mu}_e$, and the $D/B$ ratio]{Distribution of observed 
photometric parameters: 
$\log{d_n}$, $\log{r_e}$ ($d_n$ and $r_e$ in arcsec), 
$\bar{\mu}_e$, and the $D/B$ ratio.}
\label{fig:histphot}
\end{figure}

All images were reduced using standard {\it IRAF} tools, and surface
photometry was carried out by fitting elliptical and circular aperture
to the two-dimensional light distribution of each galaxy.  Light
profiles were fitted either by a pure de Vaucouleurs' r$^{1/4}$ law or
in combination with an exponential profile to represent the light
distribution of bulge+disk systems. The fits used the algorithm which
was developed in the EFAR survey (Saglia \etal 1997). This also
corrects for the smearing effects due to seeing and provides
information about the quality of the fit. The main photometric
parameters derived are $d_n$, the angular diameter within which the
mean surface brightness is equal to $\bar {\mu}_R = 19.25$ mag
arcsec$^{-2}$, the half-light radius, $r_e$, the mean surface
brightness within this radius, \mue, as well as global quantities such
as total magnitudes, $m_R$, the disk-to-bulge ratio, $D/B$,
ellipticity, position angle and parameters which can be used to
characterize the shape of images.  Figure~\ref{fig:histphot} shows the
distribution of $\log d_n$, $\log{r_e}$, \mue and the disk-to-bulge
ratio $D/B$ for all observed galaxies for which these parameters could
be measured. The first three parameters are used in defining the
distance relations, while $D/B$ has been used as an indicator of
possible misclassifications or inadequate light profile fitting.
Details of the observations, data reduction, analysis and tables
listing the photometric parameters will be presented by Alonso \etal
(2000 a,b).

As mentioned above our derived values of photometric parameters
($\log~d_n$, $\log~r_e$, \mue) were transformed into an internally
consistent system using repeated observations. In total, 498
observations of 358 galaxies are available for assessing these
statistical corrections. By careful examination of all multiple
observations available, internal conversion relations were derived for
each photometric parameter of interest and were used to create a
uniform internal system.  In addition, we observed 346 galaxies in
common with other sources, including 100 galaxies in the 7S sample for
which no images were previously available, to derive corrections to
scale the data in the public domain to our system.  From the
comparison between our measurements and those of other authors we
estimate that the typical errors in $\log d_n$, $\log{r_e}$, and \mue
are 0.017~dex, 0.08~dex, and 0.3~\bri, respectively. These values are
consistent with our internal estimates. Finally, it is also worth
mentioning that some $\sim$150 galaxies have also been observed in the
$B$-band and $\sim$200 in the $J$ and $K^\prime$ near-IR bands. These
data will be presented in forthcoming papers.

\subsection{ENEAR: The Catalog of Early-type Galaxies}

In the course of this work we have assembled all of the spectroscopic
and photometric data gathered by the ENEAR survey, the spectroscopic
data from other ongoing observations in the Southern hemisphere and
most of the available data in the literature into a uniform catalog of
early-type galaxies, which we will refer to as the ENEAR catalog. The
main goal of producing this catalog was to minimize the number of
galaxies that required observations, and to produce a homogeneous
database for other applications.  Even though extensive, the
compilation from the literature is not complete as preference was
given to data in the redshift range of interest. Table~\ref{tab:ref}
summarizes the contribution from different sources included in the
present catalog listing: in column (1) the source; in column (2) the
number of galaxies with measured central velocity dispersion; in
column (3) number of galaxies with measured \mg line index; in column
(4) the number of galaxies with measured $d_n$; and in column (5)
number of galaxies with available FP parameters. Also included are the
older compilations of central velocity dispersion measurements carried
out by Tonry \& Davis (1981) and Whitmore \etal (1985), even though,
in general, these data are not used in combination with newer data. We
also give the number of early-type galaxies measured by our group in
the Southern hemisphere which are not part of the ENEAR survey, as
described in this paper, but have been already analyzed by Rit\'e
(1999). We refer to this additional sample as ENEAR+.

From the literature, the bulk of the data still comes from the all-sky
sample of 7S galaxies (\eg Faber \etal 1989 and as distributed by
D. Burstein in electronic form, the so-called Mark~II catalog). A
revised version of the spectroscopic parameters has recently been
kindly provided by D.  Burstein.  With the exception of this sample,
most other available data sets are primarily in clusters of galaxies.
It is also important to emphasize that the data as a whole are very
inhomogeneous.  The spectroscopic parameters were derived from spectra
obtained with very different spectrographs (\eg photon-counting
detectors, using small slits; CCD detectors, using long-slits). The
original spectra also have a wide range of resolutions and spectral
coverage.  Similarly, the photometric data include old photo-electric
photometry and images using small and large format CCDs, and were
obtained using different filters.

Therefore, the measured parameters cannot be used as listed in the
original sources but must be first transformed into a common scale,
utilizing the conversion relations derived from overlapping data as
described above. In our final catalogs whenever data from other
authors are used they have been corrected as described in Bernardi
\etal (1999a).  This procedure has been helpful in maximizing the use
of publicly available data and has enabled us to reach a higher
completeness of the ENEAR samples.

Currently, in our database there are $\sim 4500$ measurements of
velocity dispersion, 3700 values of the $d_n$ parameter and 2304
values of the combination $FP = \log~r_e$ - 0.3\mue, as used in the FP
relation. These yield 2797 galaxies with measured central velocity
dispersion, 1960 galaxies with $d_n$, 1588 galaxies with $FP$, 1847
and 1458 galaxies with the required information to compute their
distances using the \dnsig and FP relations, respectively.  From these
data, distances to 76\% of the ENEARm galaxies and 89\% of the ENEARc
sample can be estimated.

In our final catalog, galaxies with more than one measurement of
$d_n$, redshift and velocity dispersion were combined, discarding
measurements from compilations, using an error-weighted mean of the
different measurements, eliminating outliers whenever possible.

\section {Redshift-distance Catalog for Peculiar Velocity Analysis}
\label{redshiftdistance}

\subsection{Distance Relation}
\label{distance}

Measurement of the radial component of galaxy's peculiar velocity,
$v_p=cz-R$, requires knowing its redshift $cz$ and its redshift
independent distance $R$ (in units of \kms), which must be determined
using a secondary distance indicator. Distances to early-type galaxies
may be inferred from relations such as \dnsig and FP, which relate
different observable quantities that depend on the distance of the galaxy (\eg
$d_n$, $FP$) to a redshift independent quantity such as the central
velocity dispersion. For studies of the peculiar velocity field it is
common practice to derive these relations for galaxies in clusters,
presumed to be at the same distance.  To improve the statistics data
from different clusters are combined to derive a template relation
(\eg Giovanelli \etal 1997b, Bernardi \etal 1999b).

For the purpose of deriving \dnsig and FP template relations we have
compiled the ENEARc sample described above. It comprises 577 galaxies
with $d_n$, 473 with $FP$ parameters, and 586 with $\sigma$, yielding
a sample of 569 and 431 cluster galaxies to derive a \dnsig and FP
relations, respectively. Of these we have obtained new $R-$band
photometric parameters for 315 galaxies, high-quality spectra for 235,
and combined with other available data we have added 153 new distance
measurements, which 93 had no prior measurements.

Until recently most studies of peculiar velocity using early-type
galaxies relied on \dnsig distances, with the FP being used primarily
for more detailed stellar population studies and of possible
environment effects. While it is our intention to compute both
distance relations and investigate possible effects on the results of
the analysis of the peculiar velocity field, our preliminary analysis
of the data is based on the \dnsig relation derived by Bernardi \etal
(1999b), postponing to a future paper the derivation of the FP
template relation (Alonso \etal 2000c). This choice was motivated by
the larger number of measurements of $d_n$ available, the fact that
this quantity is less sensitive to seeing effects, and the
parameterization used in fitting the light profiles, as required for
determining the half-light radius $r_e$ and the mean surface
brightness \mue that enter in the FP relation. Moreover, recent
studies have shown that the scatter about this relation is comparable
to that obtained using the FP relation (\eg J{\o}rgensen, Franx \&
Kj{\ae}rgaard 1996; D'Onofrio \etal 1997).

\subsection {Pruning the sample}
\label{pruning}

As described in section~\ref{selection}, The ENEARm sample is defined
from a combination of catalogs, therefore, it is not surprising that
some pruning of the sample is required both before and after the
observations. In general, galaxies were inspected before the
observations, especially after the availability of the DSS. At this
point some galaxies were immediately removed because of one of the
following reasons: stellar contamination; crowded fields preventing a
good determination of the background; interacting galaxies;
contamination/superposition by neighboring galaxies of comparable size
and/or brightness; and morphological misclassifications, especially
late-type spirals.  We have also discarded galaxies during the
analysis of CCD images whenever they revealed features not discernible
in the DSS due to saturation in the inner parts of the images. A
common cause is the presence of relatively bright stars superposed on
the central parts of the galaxy image, making the reliable
determination of photometric parameters impossible. A complete
description of these cases will be presented in Alonso \etal (2000a).
We have also removed $\sim$ 100 galaxies from the sample after
analyzing their spectra. These cases include galaxies with strong
emission-lines but very weak absorption features, galaxies with small
velocity dispersions (depending on the spectral resolution used) and
low surface brightness galaxies with low $S/N$ spectra ($\lsim 10-15$)
(see Wegner \etal 2000). In total, roughly 250 galaxies were discarded
for these reasons reducing the ENEARm sample from 1847 to 1607. This
culled sample is used below as a reference in estimating the
completeness of the ENEAR survey.

Finally, when compiling the peculiar velocity catalog, galaxies with
extremely large peculiar velocities were re-examined and those with old
photometric or spectroscopic measurements or with some peculiarity in
their images (prominent bars, shells, spiral arms, dust lane) that
could either affect the derivation of the photometric parameters of
interest or lead to a significant departure of the galaxy from the
distance relation were removed ($\sim 200$ galaxies) from the sample
used for peculiar velocity analysis (Alonso \etal 2000d).

\subsection {Grouping galaxies}
\label{group}

In contrast to late-type galaxies, early-type objects tend to reside in
virialized clumps and in regions of high density. Therefore, in order
to use early-type galaxies to map the peculiar velocity field it is
important to assign them to groups and clusters, in order to minimize
the impact of virial motions on the analysis of the velocity
field. One advantage of grouping galaxies is that the distances to
grouped objects are more accurately determined, as the fractional
distance errors are reduced by a factor of $\sqrt N$, where $N$ is the
number of early-type galaxies in the group.

Since our sample has been drawn from complete redshift surveys,
membership assignment of galaxies to groups can be done in a
systematic way, using objective friends-of-friends algorithms to
search for galaxy groups in the original magnitude-limited
catalogs. To this end we have used the catalogs of groups of Maia
\etal (1989) for the SSRS and of Geller \& Huchra (1983) for the
CfA1. These groups were defined as associations of galaxies with a
density contrast $\delta \rho/ \rho > 20$.  These catalogs have been
used as they provide the largest sky coverage. In regions covered by
CfA2 and SSRS2 we have replaced these groups by those recently
compiled by Ramella \etal (1997) for the CfA2 and by Ramella \etal
(1999) for the SSRS2. These surveys cover a slightly smaller area but
extend to fainter magnitudes and consider groups with larger density
contrast $\delta \rho/\rho > 80$ which, as demonstrated by numerical
simulations, are less contaminated by spurious groups
(members). Finally, groups have also been identified at low galactic
latitudes using the ORS data kindly provided by B. Santiago and
M. Davis.  Using the position and parameters of these groups,
early-type galaxies were assigned to a group if their projected separation
relative to center of a group was $\leq 1.5R_p$ and their redshift
satisfied the condition ($cz_{\rm gal} - cz_{\rm gr}) \leq
1.5\sigma_{\rm gr}$. Here, $R_p$ is the mean projected separation of
the group (Ramella \etal 1989), $cz_{\rm gal}$ is the radial velocity
of the early-type galaxy, $cz_{\rm gr}$ is the mean group velocity and
$\sigma_{\rm gr}$ is the velocity dispersion of galaxies in the
group. We adopted these criteria in an attempt to homogenize the
assignment of galaxies to groups, given the differences in density
contrast used in their original identifications, and to allow for the
identification of systems which may have been missed near the edges of
the catalogs.

\begin{figure}
\centering
\mbox{\psfig{figure=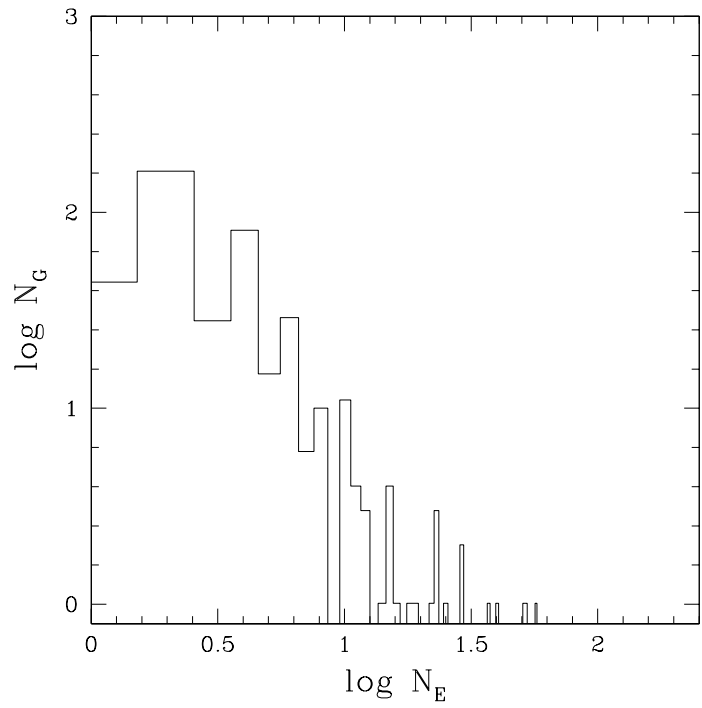,height=10truecm}}
\caption{Multiplicity function of early-type galaxies showing the number
of  groups as a function of the number of early-type galaxies in the
groups.}
\label{fig:multiplicity} 
\end{figure}

Adopting the above criteria we find that 59\% of the early-type
galaxies in ENEARm are in groups. We point out that 15\% of the
galaxies are associated with groups in which the other members are
late-type galaxies. Such systems represent about 10\% of the
total number of groups. In the latter case the peculiar velocity of
the galaxy is computed using the galaxy's distance, while the redshift
is that of the group. Groups containing two or more early-type
galaxies are treated as single objects with the redshift being given
by the mean redshift of the group, including all morphological
types, and the logarithm of the distance as the error-weighted mean of
the logarithm of the distance of the early-type galaxies in that
group. Figure~\ref{fig:multiplicity} shows the number of groups as a
function of the number of early-type galaxies in the group. We note
that in the case of clusters within $cz\leq7000$~\kms used to
determine the distance relations (see section~\ref{clusters}) we have
also used fainter galaxies in the calculation of their redshift and
distance. As discussed above (see also Bernardi \etal 1999a), we
arbitrarily define clusters to be systems with more than 10 members,
of which at least 5 are early-type galaxies.  This cluster sample is
used by Bernardi \etal (1999b) and Alonso \etal (2000c) to define the
template distance relation later used to estimate galaxy
distances. Finally, we point out that the membership assignment
criteria described above was adopted by Bernardi \etal (1998) to
investigate the properties of early-type galaxies in different
environments.

\begin{figure}
\centering
\mbox{\psfig{figure=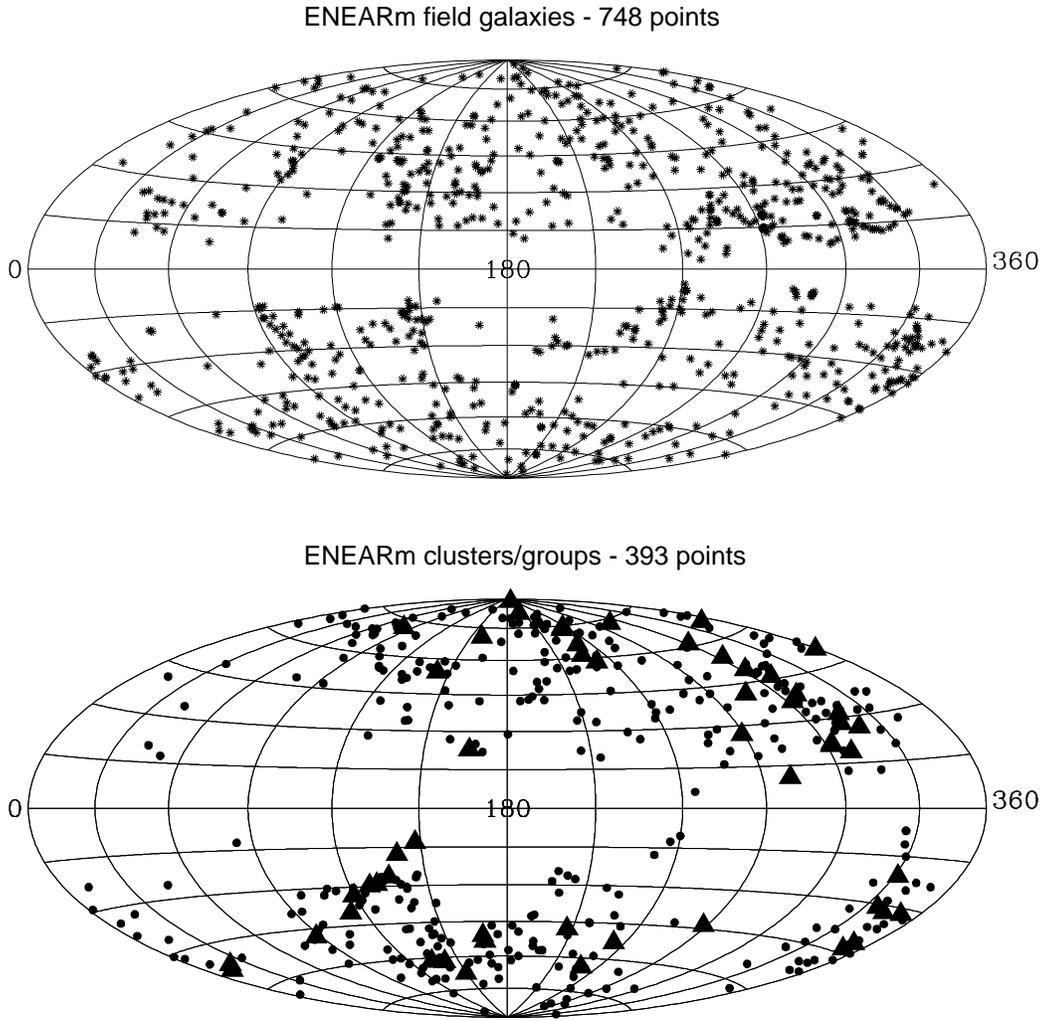,height=15truecm,bbllx=4truecm,bblly=5truecm,bburx=19truecm,bbury=25truecm }}
\caption{The projected sky distribution, in galactic coordinates, of
the ENEARm objects. The upper panel shows individual galaxies
considered as isolated by the membership criteria adopted.  The lower
panel shows the distribution of ``objects'' to which objects have been
assigned, distinguishing groups (solid circles) and clusters (solid
triangles) as defined in the text.  }
\label{fig:enearobj}
\end{figure}

Figure~\ref{fig:enearobj} shows the projected distribution of the 1141
``objects'' in the ENEARm sample. The upper panel shows the 748
galaxies not assigned to groups, while the lower panel shows 393
groups/clusters.  As expected, the ``isolated'' galaxies are more
uniformly distributed across the sky, while the group/cluster sample
delineates more clearly the most prominent large-scale structures in
the nearby universe.

\subsection{Completeness}

\begin{figure}
\centering
\mbox{\psfig{figure=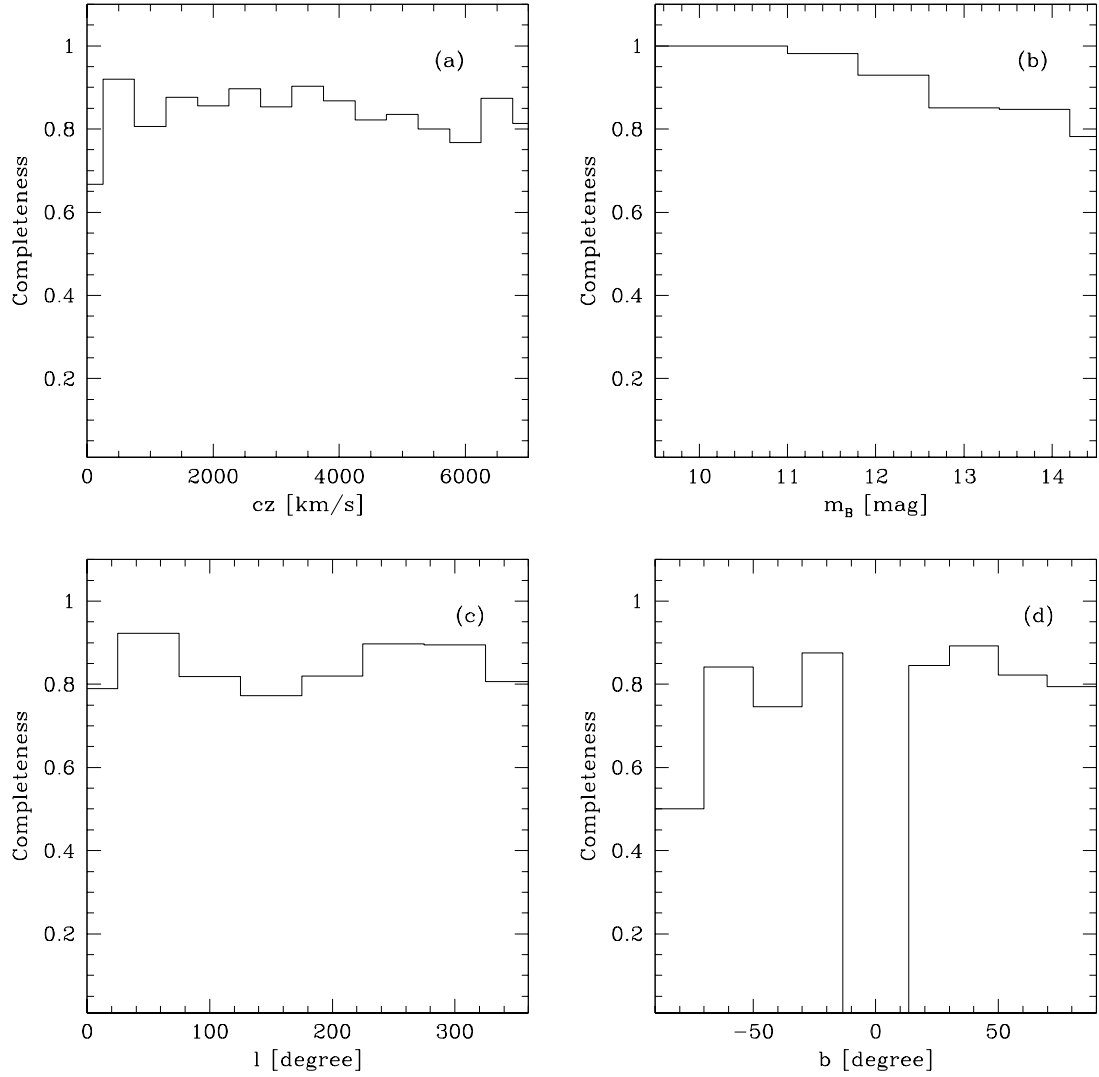,height=16truecm}}
\caption{The completeness of the ENEARm sample as function of: redshift
(panel a), magnitude, $m_B$ (panel b); galactic longitude (panel c);
and galactic latitude (panel d).}
\label{fig:complet}
\end{figure}

Since our parent sample has complete redshift information,
completeness here refers to galaxies for which we have both velocity
dispersion and $d_n$ measurements and thus \dnsig
distances. Figure~\ref{fig:complet} shows the completeness function of
the ENEARm redshift-distance survey as a function of: (a) redshift,
(b) blue magnitude $m_B$, (c) galactic longitude, and (d) galactic
latitude. The completeness is computed relative to the sub-sample of
1607 galaxies that remain after excising galaxies, from the original
sample of 1847 galaxies, unsuitable for distance estimates, as
discussed above.  Examining the figure we find no strong dependence
with the redshift, magnitude, or galactic longitude and latitude
except near the southern galactic pole ($b \lsim -70^\circ$). In
general, the sample completeness is a nearly constant $\gsim 80$\%.
This demonstrates that our dataset, even though not complete is
remarkably uniform both in depth and in sky coverage.  This is in
marked contrast to other catalogs assembled from different sources
(\eg Mark~III) which show large sampling variations in different
directions of the sky.  We have not investigated the completeness of
the sample as function of morphological types since analysis of our
images shows that the morphological classification available from the
original catalogs is unreliable and care should be exercised in
defining sub-types of the early-type population for any detailed
analysis of their properties.  We note that the information presented
here is useful for producing realistic mock samples for analysis of
the peculiar velocity field.

\begin{figure}
\centering
\mbox{\psfig{figure=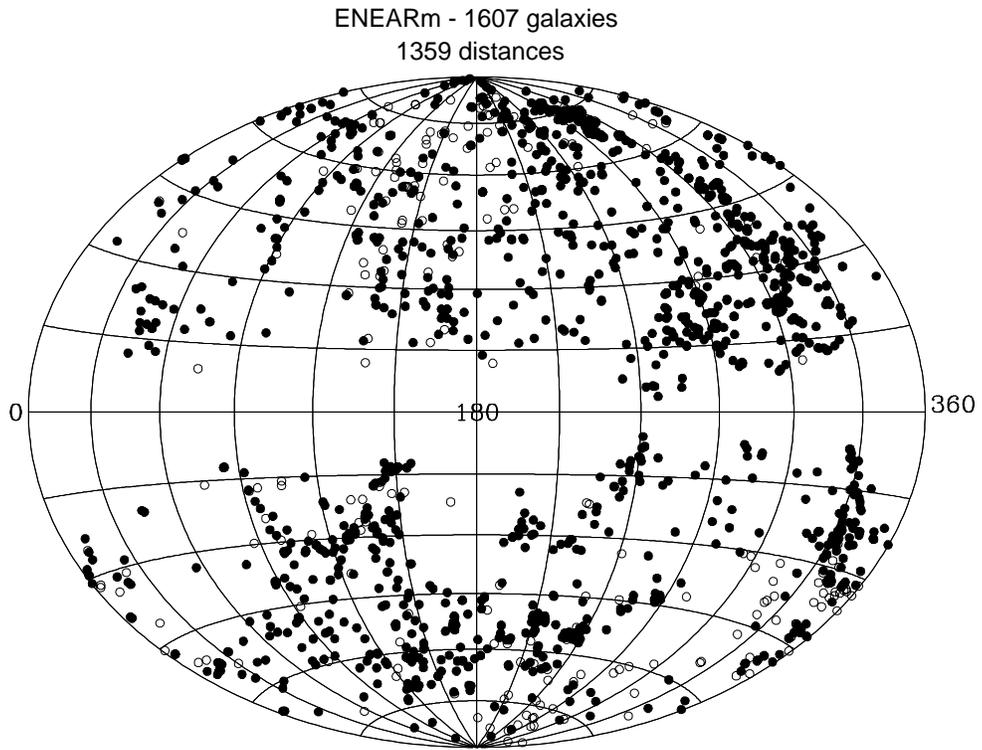,height=13truecm,bbllx=3truecm,bblly=7truecm,bburx=19truecm,bbury=25truecm}}
\caption{Projected sky distribution, in Galactic coordinates, of all
ENEARm galaxies with (filled circles) and without (open circles)
measured distances.}
\label{fig:enearfdist}
\end{figure}

To underscore the uniformity of the observed sample
Figure~\ref{fig:enearfdist} compares the projected distribution, in
galactic coordinates, of all galaxies of the ENEARm sample that have
measured distances (filled circles) with galaxies for which distances
are still unavailable (open circles).  Clearly, the sample with
measured distances delineates well all major structures in the nearby
universe. We also find that the distribution of missing galaxies does
not reveal any particularly under-sampled region.

\begin{figure}
\centering
\mbox{\psfig{figure=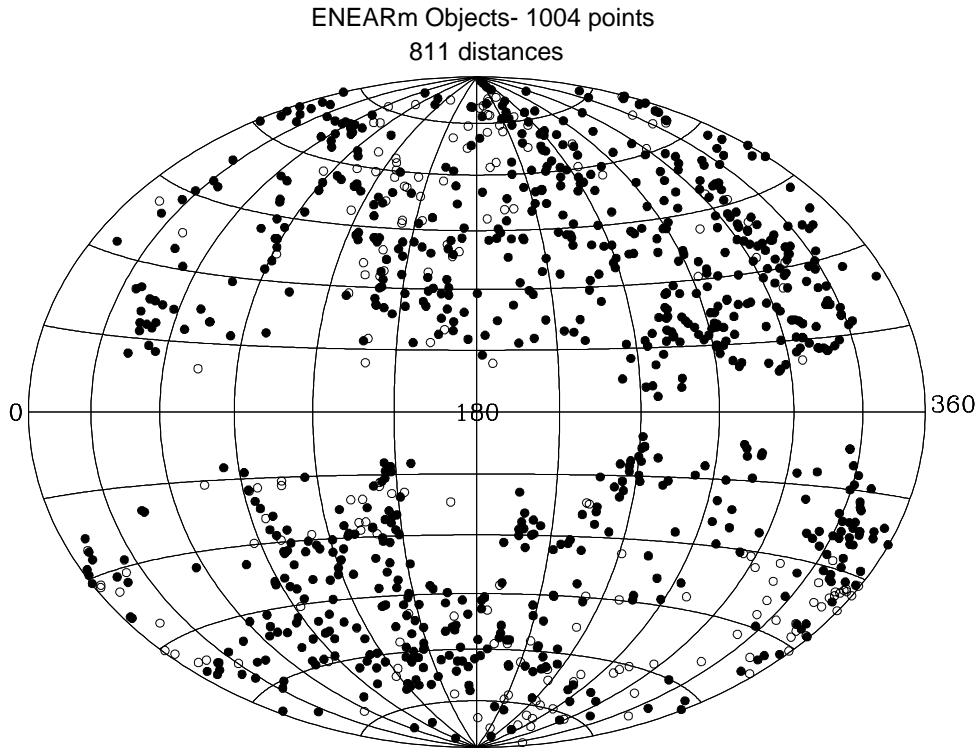,height=13truecm,bbllx=3truecm,bblly=7truecm,bburx=19truecm,bbury=25truecm}}
\caption {Projected sky distribution, in galactic coordinates, of
ENEARm ``objects'' useful for peculiar velocity analyses showing
objects with measured distances (full circles)  and
without (open circles).}
\label{fig:eneardistobj}
\end{figure}

So far, we have shown all galaxies individually. However, as discussed
earlier, early-type galaxies are found predominantly in
clusters/groups.  Therefore, Figure~\ref{fig:eneardistobj} shows the
distribution of independent objects defined by the grouping procedure
described in section~\ref{group} (and in more detailed in Bernardi
\etal 1999a).  Currently, we have distances for 1644 galaxies, 1359 in
ENEARm. The remaining are galaxies in the ENEARc cluster sample but
not in the ENEARm. We include these galaxies because they are actually
used in the definition of the redshift, distance and error estimate
for these clusters.  Figure~\ref{fig:eneardistobj} shows that there
are 811 independent objects of which 6 are clusters beyond
7000~\kms. As briefly mentioned in section~\ref{pruning} about 200
galaxies show, after closer inspection of cases with extreme values of
the peculiar velocity, features in their spectra and/or in their
images that can affect the measured parameters that enter the distance
relation and as a precaution were removed from the sample. A detailed
description and listing of these cases will be presented by Alonso
\etal (2000d). At the present time the sample being used in our
preliminary analysis (da Costa \etal 2000; Zaroubi \etal 2000)
consists of 1430 galaxies grouped in 702 objects.

\subsection{Comparison with other surveys}

\begin{figure}
\centering
\mbox{\psfig{figure=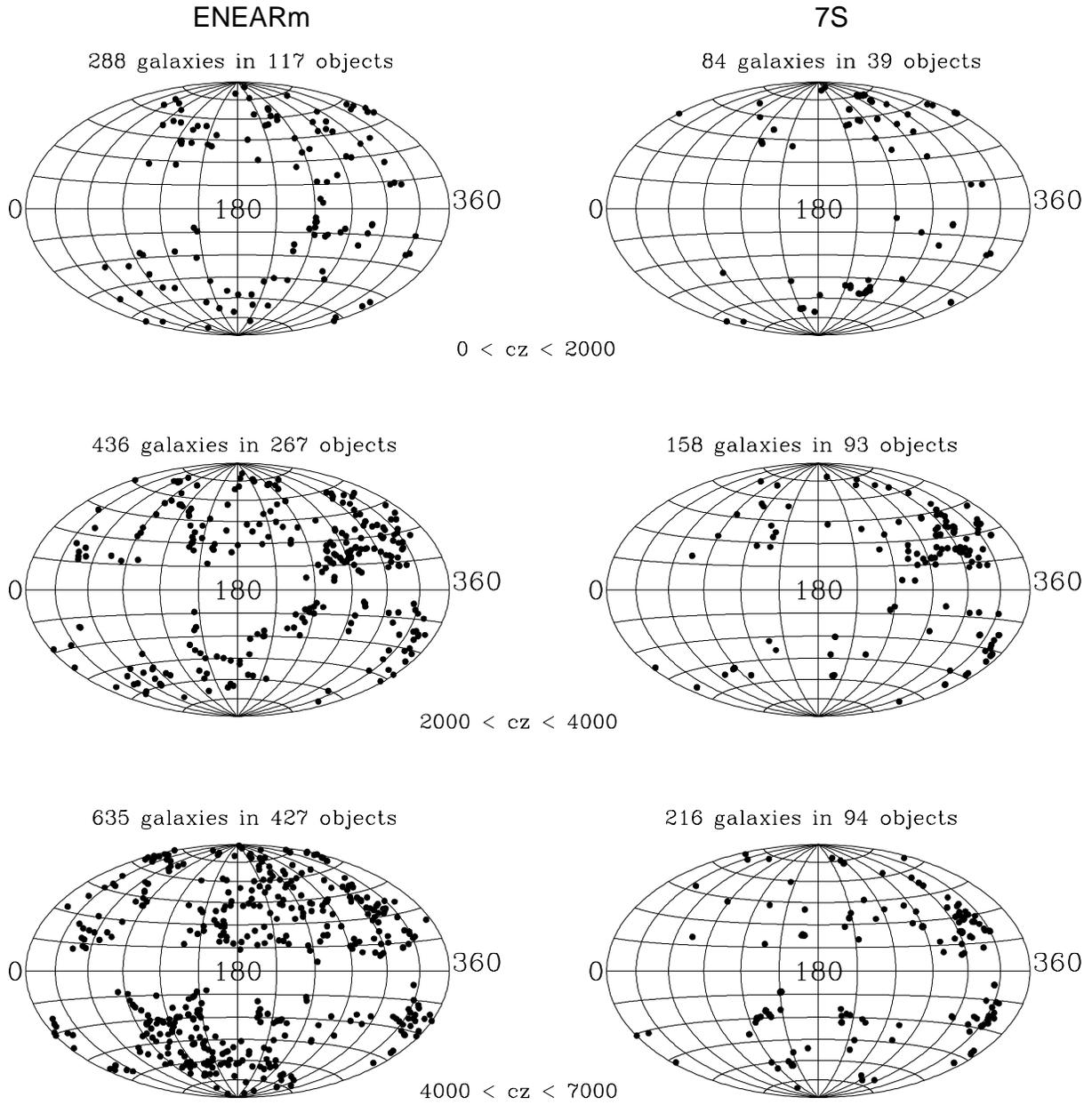,height=17truecm,bbllx=4truecm,bblly=6truecm,bburx=19truecm,bbury=25truecm }}
\caption{Comparison of the projected distribution, in galactic
coordinates, of ENEARm and 7S objects in different redshift
slices. The number of galaxies and the number of objects, after
grouping, with measured distances are given for each redshift shell.}
\label{fig:enearmark2}
\end{figure}

The ENEARm sample shown in Figure~\ref{fig:enearfdist} is the largest
and most homogeneous sample of nearby early-type galaxies currently
available for cosmic flow studies. To highlight this,
Figure~\ref{fig:enearmark2} compares the projected distributions, in
redshift slices, of the ENEARm and the 7S samples after galaxies are
grouped.  Besides the obvious differences in the total number of
galaxies and objects, there is a striking difference in the structures
probed by the two samples.  In particular, in the most distant
redshift shell, Perseus-Pisces is clearly visible in our catalog while
is conspicuously absent in the 7S sample.  This accounts for some of
the surprises encountered in earlier (7S-based) reconstructions (Dekel
\etal 1990).

\begin{figure}
\centering
\mbox{\psfig{figure=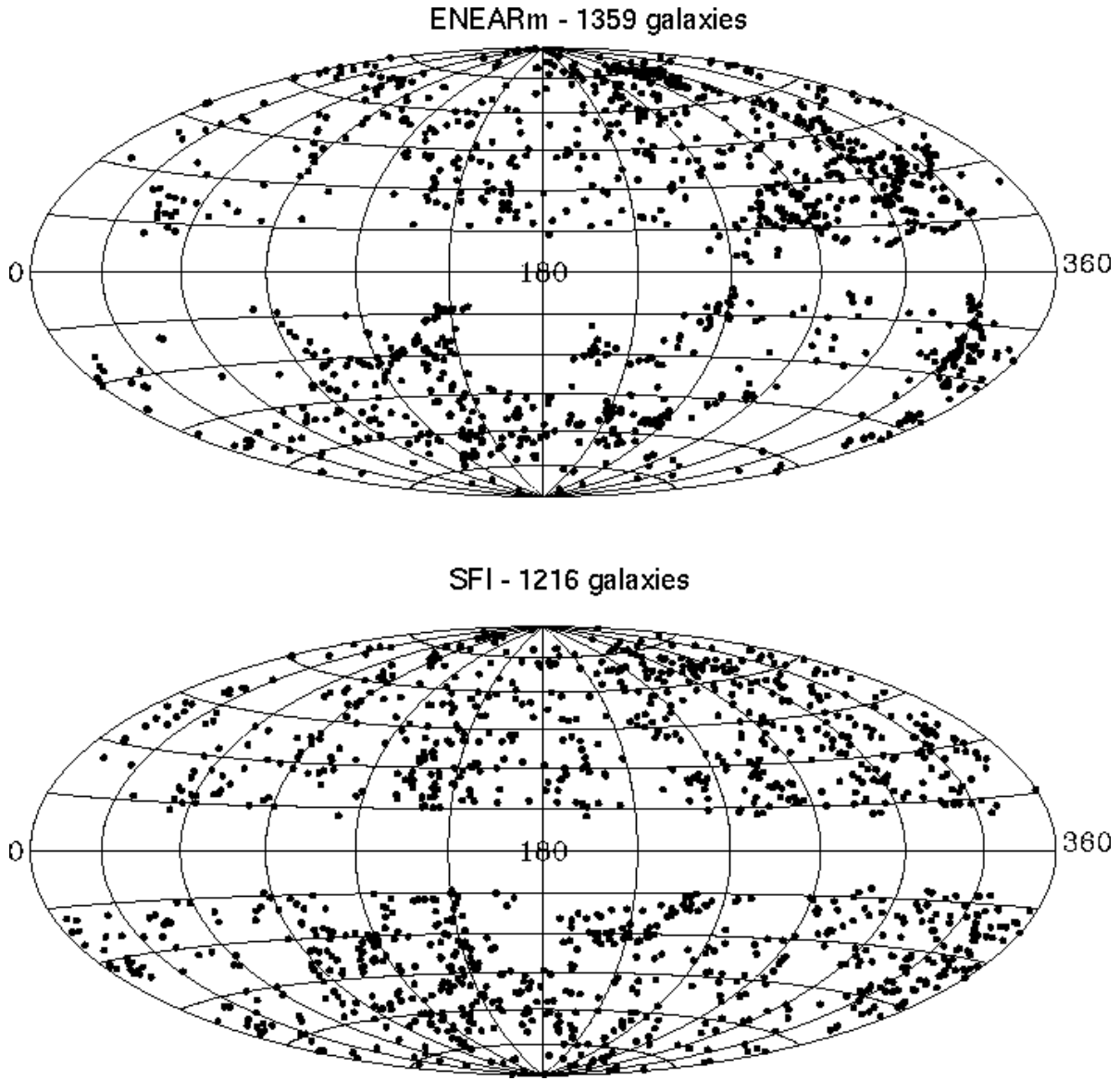,height=15truecm,bbllx=4truecm,bblly=5truecm,bburx=19truecm,bbury=22truecm }}
\caption{Comparison of the
projected distribution of individual galaxies, in galactic
coordinates, from the ENEARm sample (upper panel) and the SFI catalog
of spiral galaxies (lower panel).}
\label{fig:enearmark2sfi}
\end{figure}

It is also worth comparing our sample with catalogs built from
recently completed TF surveys.  Figure~\ref{fig:enearmark2sfi}
compares the projected distribution of individual galaxies having
measured peculiar velocities in ENEARm and the SFI TF-survey of spiral
galaxies. From the figure one can see the complementarity of these
samples. The ENEARm early-type galaxies delineate the structures more
sharply while the spirals are clearly more uniformly distributed. For
this reason, the combination of the two samples is highly desirable,
and it is the subject of future work.

\begin{figure}
\centering
\mbox{\psfig{figure=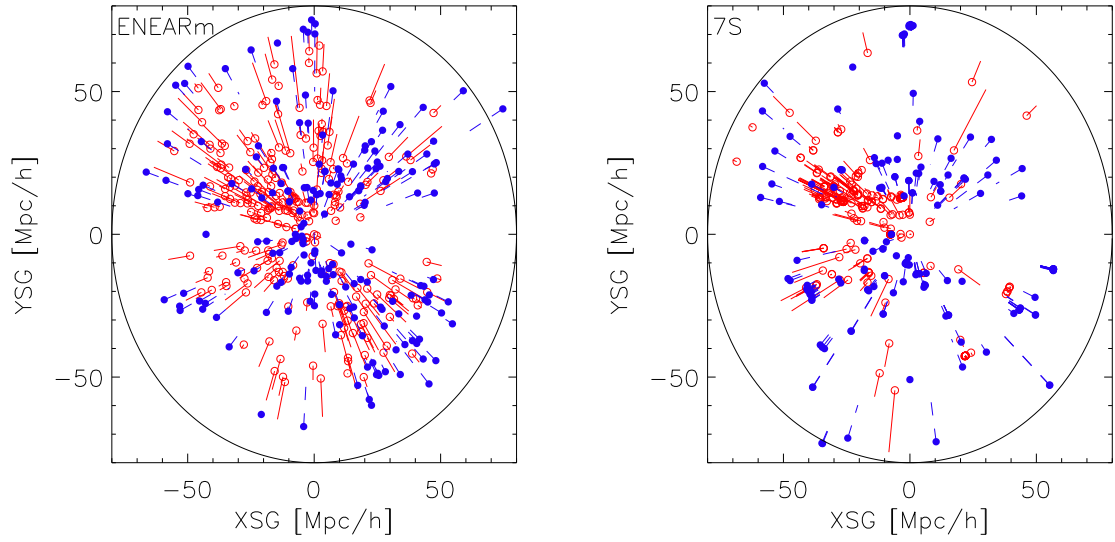,height=15truecm,bbllx=4truecm,bblly=5truecm,bburx=19truecm,bbury=22truecm }}
\caption{Map of the measured radial component of the peculiar velocity
field in the CMB restframe along the Supergalactic Plane for the
ENEARm (left panel) and 7S (right panel) objects within 5000\kms of
this plane. Full symbols depict infalling galaxies and open symbols
with dashed extensions outflowing galaxies.}
\label{fig:SGslice}
\end{figure}

Finally, as a preview for future papers in this series we present in
Figure~\ref{fig:SGslice} the homogeneous Malmquist bias corrected
peculiar velocities of the ENEARm galaxies within $\pm 5000$~\kms from
the Supergalactic plane in the CMB rest frame. For comparison we show
the map of the flowfield as probed by the 7S. The most remarkable
features are the suggestion of a back-side infall in the direction of
the Great Attractor region and the flow associated with the
Perseus-Pisces complex nearly absent in the 7S sample. A detailed
analysis of the velocity field will be presented in subsequent papers
of this series.

\section {Summary}
\label{summary}

We have presented an overview of the ongoing imaging/spectroscopic
survey of nearby early-types being conducted in both hemispheres which
allowed the construction of an extensive and homogeneous database
consisting of redshifts, central velocity dispersion measurements,
spectral line indices and photometric data for roughly 3400 early-type
galaxies, with multiple measurements in the nearby Universe
($z\lsim0.05$). Since these galaxies are part of complete redshift
surveys additional information is also available regarding the type of
environment and measures of the local galaxy density, making the
catalog useful for different applications.

We have also discussed in some detail how a magnitude and
redshift-limited sample of galaxies with measured distances and
redshifts suitable for studies of the local peculiar velocity field
has been constructed. The resulting ENEARm sample together with the
cluster galaxy sample ENEARc, constitute the basis of the work being
carried out by our  group to map the peculiar velocity field as probed
by early-type galaxies. We hope that the comparison of the peculiar
velocity fields derived from the ENEAR survey and similar TF-based
surveys will provide a more definite answer about the true physical
nature of the observed motions, thus supporting the assumption that
the peculiar motions observed are in fact induced by the gravity field
associated to fluctuations of the underlying mass density field.

Quantitative analysis of various aspects of the measured velocity field
such as the calculation of the dipole, the mass power-spectrum, and
determinations of $\beta$ will be presented in subsequent papers of this
series. In the future, we also hope to be able to combine the ENEAR and
SFI samples to obtain improved maps of the underlying mass distribution.
 Finally, we also anticipate the use of our dataset for more detailed
stellar population studies.

We point out that observations are still underway, with the aim of
completing the ENEARm sample and to extend the number of clusters
available for the construction of distance relations.

\acknowledgments{The authors would like to thank all of those who have
contributed directly or indirectly to this long-term project too many
to enumerate individually.  Our special thanks to Ot\'avio Chaves for
his many contributions over the years and for conducting several of
the observations reported here. We would also like to thank
D. Burstein. M. Davis, A. Milone, M. Ramella, R. Saglia, B. Santiago
and S. Zaroubi for useful discussions and input.  LNdC would like to
extend his special thanks to David W. Latham who played a pivotal role
at the early stages of this project and the support received from the
CNPq, the Smithsonian Astrophysical Observatory, the Guggenheim
Foundation, the Institut d' Astrophysique (IAP) and the Racah
Institute for Physics of the Hebrew University, in different phases of
this project.  MB thanks the Sternwarte M\"unchen, the Technische
Universit\"at M\"unchen, ESO Studentship program, and MPA Garching for
their financial support during different phases of this research. MB
also acknowledges travel support provided by the ESO Science Division
and the Observat\`orio Nacional. MVA thanks CNPq for
different fellowships at the beginning of the project and the CfA and
ESO's visitor programs for support of visits.  MVA is partially
supported by CONICET, SecyT and the Antorchas--Andes-- Vitae
cooperation. GW is grateful to the Alexander von Humboldt-Stiftung for
making possible a year's stay at the Ruhr-Universit\"at in Bochum, and
to ESO for support for visits to Garching which greatly aided this
project.  Financial support for this work has been given through
FAPERJ (CNAW, MAGM, PSSP), CNPq grants 201036/90.8, 301364/86-9
(CNAW), 301366/86-1 (MAGM); NSF AST 9529098 (CNAW); ESO Visitor grant
(CNAW). PSP and MAGM thank CLAF for financial support and CNPq
fellowships. Most of the observations carried out at ESO's 1.52m
telescope at La Silla were conducted under the auspices of the
bi-lateral time-sharing agreement between ESO and MCT/Observat\'orio
Nacional. This research has made use of the NASA/IPAC Extragalactic
Database (NED) which is operated by the Jet Propulsion Laboratory,
CALTECH, under contract with the National Aeronautics and Space
Administration. We acknowledge the use of NASA's $SkyView$ facility
(http://skyview.gsfc.nasa.gov) located at NASA Goddard Space Flight
Center and of the digitized sky survey, produced at the Space
Telescope Science Institute under U.S. Government grant NAG
W-2166. The images are based on photographic data obtained using UK
Schmidt Telescope, operated by the Royal Observatory Edinburgh, with
funding from the UK Science and Engineering Research Council (later
the UK Particle Physics and Astronomy Research Council), until 1988
June, and thereafter by the Anglo-Australian Observatory. Part of the
observations carried out at CASLEO made use of the CCD and data
acquistion system supported under U.S.  National Science Foundation
grant AST-90-15827 to R.M. Rich.}

{}

\newpage

\clearpage

\clearpage

\clearpage
\clearpage

\clearpage

\clearpage

\clearpage

\clearpage

\clearpage

\clearpage

\clearpage

\clearpage

\clearpage

\newpage


\begin{table}
\begin{center}
\title {Samples in the ENEAR catalog}
\label{tab:ref}
\begin{tabular}{lrrrr}
\\
\tableline \tableline
Source   &   N$_{\sigma}$ & N$_{{\bf Mg}_2}$ & N$_{D_n}$ & N$_{{\bf
FP}}$ \\
 (1) & (2) & (3) & (4) & (5)\\
\tableline
ENEAR            & 1210 & 1036 & 1398 & 1398 \\
ENEAR+           &  792&  792 & --  & --- \\
TD (1981)               & 229 & --  & -- & -- \\
WMT (1985)              & 130 & --  & -- & -- \\
LC (1988)             & 114 & --  & 78  & -- \\
7S (1989)            & 543 & 533 & 499 & 179\\    
D  (1991)            & 158 & 151 & 175 & -- \\
JFK (1992, 1995)            & 159 & 119 & 200 & 194\\
Lc  (1997)           & 85  &  -- & 84  & -- \\ 
S   (1997)           & 88  &  81 & 98  &  98 \\
\tableline
\end{tabular}
\end{center}
\tablecomments{In the second column we give only the number
of galaxies we have included in our catalog from the different
sources.  The original papers may contain more galaxies than listed
here.}
\tablerefs{
TD: Tonry \& Davis (1981); 
WMT: Whitemore, McElroy \& Tonry (1985); 
LC: Lucey \& Carter (1988);
7S: Faber \etal (1989);
D: Dressler (1987), Dressler \etal (1991);
JFK: J{\o}rgensen, Franx \& Kj\/aergaard (1992, 1995a, 1995b);
Lc: Lucey \etal (1997);
S: Smith \etal (1997).}
\end{table}


\newpage

\begin{thebibliography}{}

\bibitem []{} Aaronson, M., Huchra, J., Mould, J., Schechter, P. L., \&
Tully, R. B. 1982, ApJ, 258, 64
\bibitem []{} Alonso, M. V., da Costa, L. N., Pellegrini, P. S. \&
Kurtz, M.  J., 1993, AJ, 106, 676
\bibitem []{} Alonso, M. V., da Costa, L. N., Latham, D. W.,
 Pellegrini, P. S., Milone, A. A. E, 1994, AJ, 108, 1987
\bibitem []{} Alonso, M. V., Bernardi, M., Wegner, G. \etal 2000a, in preparation
\bibitem []{} Alonso, M. V., Bernardi, M., Wegner, G.\etal 2000b, in preparation
\bibitem []{} Alonso, M. V., Bernardi, M., da Costa, L. \etal 2000c, in preparation
\bibitem []{} Alonso, M. V., Bernardi, M., da Costa, L.\etal 2000d, in preparation
\bibitem []{} Bernardi, M., Renzini, A., da Costa, L. N., Wegner, G.,
Alonso, M. V., Pellegrini, P. S., Rit\'e, C. N. A., \& Willmer,
C. N. A. 1998, ApJ, 508, L143
\bibitem []{} Bernardi, M. \etal 1999a, in preparation
\bibitem []{} Bernardi, M. \etal 1999b, in preparation
\bibitem []{} Blakeslee, J. P., Davis, M., Tonry, J. L., Ajhar, E. A., \&
Dressler, A. 1999, in Cosmic Flows: Towards and Understanding of
Large-Scale Structure, eds. S. Courteau, M. A. Strauss, \&
J. A. Willick, ASP Conf. Ser., astro-ph/9910342
\bibitem []{} Borgani, S., da Costa, L. N., Zehavi, I., Giovanelli, R.,
Haynes, M. P., Freudling, W., Wegner, G., \& Salzer, J. J. 1999,
astro-ph/9908155
\bibitem[]{} Colless, M., Burstein, D., Davies, R. L., McMahan, R. K. J.,
 Saglia, R. P., \& Wegner, G. 1999, MNRAS, 303, 813
\bibitem []{} Courteau, S., Faber, S. M., Dressler, A., \& Willick,
J. A. 1993, ApJ, 412, L51
\bibitem []{} Courteau, S., Willick, J. A., Strauss, M. A., Schlegel,
D., \& Postman, M. 1999, in Cosmic Flows: Towards and Understanding of
Large-Scale Structure, eds. S. Courteau, M. A. Strauss, \&
J. A. Willick, ASP Conf. Ser., astro-ph/9909385 
\bibitem []{} da Costa, L. N., Freudling, W., Wegner, G., Giovanelli, R.,
Haynes, M. P., \& Salzer, J. J. 1996, ApJ, 468, L5
\bibitem []{} da Costa, L. N., Nusser, A., Freudling, W., Wegner, G.,
Giovanelli, R., Haynes, M. P., Salzer, J. J., \& Wegner, G. 1998a,
MNRAS, 299, 425
\bibitem []{} da Costa, L. N., Pellegrini, P. S., Davis, M., Meiksin, A.,
Sargent, W. L. W., \& Tonry, J. L. 1991, ApJS, 75, 935
\bibitem []{} da Costa, L. N., Pellegrini, P. S., Sargent, W. L. W., Tonry, J.,
Davis, M., Meiksin, A., Latham, D. W., Menzies, J. W., \& Coulson,
I. A. 1988, ApJ, 327, 544 
\bibitem []{} da Costa, L. N., Willmer, C. N. A., Pellegrini, P. S.,
Chaves, O. L., Rit\'e, C., Maia, M. A. G., Geller, M. J., Latham,
D. W., Kurtz, M. J., Huchra, J. P., Ramella, M., Fairall, A. P.,
Smith, C., \& L\'ipari, S. 1998b, AJ, 116, 1
\bibitem []{} Dekel, A. 1994, ARAA, 32, 371
\bibitem []{} Dekel, A., Bertschinger, E., \& Faber, S. M. 1990, ApJ,
364, 349
\bibitem []{} D'Onofrio, M., Capaccioli, M., Zaggia, S. R., \& Caon,
N. 1997, MNRAS, 289, 847
\bibitem []{} Dressler, A. 1987, ApJ, 317, 1
\bibitem []{} Faber, S. M., Wegner, G., Burstein, D., Davies, R. L.,
      Dressler, A., Lynden-Bell, D., \& Terlevich, R. J. 1989, ApJS,
      69, 763
\bibitem []{} Falco \etal  ~1999, PASP, 111, 438
\bibitem []{} Freudling, W., Zehavi, I., da Costa, L. N., Dekel, A.,
Eldar, A., Giovanelli, R., Haynes, M. P., Salzer, J. J., Wegner, G., \&
Zaroubi, S. 1999, ApJ, 523, 1
\bibitem []{} Geller, M. J., \& Huchra, J. P. 1983, ApJS, 52, 61
\bibitem []{} Geller, M. J., \& Huchra, J. P. 1989, in Large Scale
Structure and Motions in the Universe, ed. M. Mezzetti (Dordrecht:
Kluwer Academic Publishers), p. 3
\bibitem []{} Giovanelli, R., \& Haynes, M. P., Freudling, W., da Costa,
L. N., Salzer, J. J., \& Wegner, G. 1998, ApJ, 505, L91
\bibitem []{} Giovanelli, R., Haynes, M. P., Herter, T., Vogt, N. P.,
 da Costa, L. N., Freudling, W., Salzer, J. J. and Wegner, G. 1997a,
 AJ, 113, 22 
\bibitem []{} Giovanelli, R., Haynes, M. P., Herter, T., Vogt, N. P.,
da Costa, L. N., Freudling, W., Salzer, J. J. and Wegner, G. 1997b,
AJ, 113, 53 
\bibitem []{} G\'orski, K. M., Banday, A. J., Bennett, C. L., Hinshaw,
G., Kogut, A., Smoot, G. F., \& Wright, E. L. 1996, ApJ, 464, L11 
\bibitem []{} Guzm{\'a}n, R., \& Lucey, J. R. 1993, MNRAS, 263, 47
\bibitem []{} Haynes, M. P., Giovanelli, R., Salzer, J. J., Wegner, G.,
Freudling, W., da Costa, L. N., Herter, T., \& Vogt, N. P. 1999a, AJ,
117, 1668
\bibitem []{} Haynes, M. P., Giovanelli, R., Chamaraux, P., da Costa,
L. N., Freudling, W., Salzer, J. J., \&  Wegner, G. 1999b, AJ, 117, 2039 
\bibitem []{} Huchra, J. P., Davis, M., Latham, D., \& Tonry, J. 1983, ApJS,
52, 89
\bibitem []{} Huchra, J. P., Latham, D. W., da Costa, L. N.,
Pellegrini, P. S., \& Willmer, C. N. A.  1993, AJ, 105, 1637
\bibitem []{} Hudson, M. J., Lucey, J. R., Smith, R. J., \& Steel, J. 1997,
MNRAS, 291, 488
\bibitem []{} Hudson, M. J., Smith, R. J., Lucey, J. R., Schlegel,
D. J., \& Davies, R. L. 1999, ApJ, 512, L79
\bibitem []{} J{\o}rgensen, I., Franx, M., \& Kj{\ae}rgaard, P. 1992,
A\&AS, 95, 489
\bibitem []{} J{\o}rgensen, I., Franx, M., \& Kj{\ae}rgaard, P. 1993,
ApJ, 411, 34
\bibitem []{} J{\o}rgensen, I., Franx, M., \& Kj{\ae}rgaard, P. 1995a, MNRAS, 273, 1097
\bibitem []{} J{\o}rgensen, I., Franx, M., \& Kj{\ae}rgaard, P. 1995b,
MNRAS, 276, 1341
\bibitem []{} J{\o}rgensen, I., Franx, M., \& Kj{\ae}rgaard, P. 1996,
MNRAS, 280, 167
\bibitem []{} Lasker, B. M., Sturch, C. R., McLean, B. J., Russell,
J. L., Jenker, H., \& Shara, M. M. 1990, AJ, 99, 2019 
\bibitem []{} Lauberts, A. 1982, The ESO-Uppsala Survey of the ESO(B)
Atlas (M\"unchen: ESO)
\bibitem []{} Lauberts, A., Valentijn, E. A., 1989, The Surface Photometry
    Catalogue of the ESO-Uppsala Galaxies (Garching: ESO)
\bibitem []{} Lynden-Bell, D., Faber, S. M., Burstein, D., Davies, R. L.,
Dressler, A., Terlevich, R., \& Wegner, G. 1988, ApJ, 326, 19
\bibitem []{} Maia, M. A. G., da Costa, L. N., \& Latham, D. W. 1989,
ApJS, 69, 809
\bibitem []{} Marzke, R. O., da Costa, L. N., Pellegrini, P. S.,
Willmer, C. N. A., \& Geller, M. J. 1998, ApJ, 503, 617
\bibitem []{} Mathewson, D. S., Ford, V. L., \& Buchhorn, M. 1992, ApJS,
81, 413 (MFB)
\bibitem []{} Mathewson, D. S., \& Ford, V. L. 1996, ApJS, 107, 97
\bibitem []{} M\"uller, K., Wegner, G. \& Freudling, W. 1999, A\&A Suppl,
in press, astro-ph/9910392
\bibitem []{} M\"uller, K., Freudling, W., Watkins, R, \& Wegner,
G., 1998, ApJL, 507, 105
\bibitem []{} Nilson, P. 1973, Uppsala General Catalog of Galaxies,
Uppsala Astron. Obs. Ann., 6 (UGC) 
\bibitem []{} Nusser, A., Davis, M., \& Willick, J. A. 1997, in Galaxy
Scaling Relations: Origins, Evolution and Applications, eds. L. da
Costa, \& A. Renzini (Berlin: Springer), p. 286
\bibitem []{} Pellegrini, P. S., da Costa, L. N., Willmer, C. N. A.,
Huchra, J. P., \& Latham, D. W. 1990, AJ, 99, 751
\bibitem []{} Ramella, M., Geller, M. J., \& Huchra, J. P. 1989, ApJ, 344,
57
\bibitem []{} Ramella, M., Pisani, A., \& Geller, M. J. 1997, AJ, 113,
483
\bibitem []{} Ramella, M., \etal ~1999, in preparation
\bibitem []{} Rit\'e, C. 1999, Ph.D. Thesis, CNPq/Observat\'orio
Nacional
\bibitem []{} Rit\'e, C., Maia, M. A. G., Willmer, C. N. A. \etal 2000, in preparation
\bibitem []{} Riess, A. G. 1999, in Cosmic Flows: Towards and Understanding of
Large-Scale Structure, eds. S. Courteau, M. A. Strauss, \&
J. A. Willick, ASP Conf. Ser., astro-ph/9908237
\bibitem []{} Saglia, R. P., Bertschinger, E., Baggley, G., Burstein, D.,
Colles, M., Davies, R. L., McMahan,  R. K., \& Wegner, G. 1997,
ApJS 109, 79
\bibitem []{} Santiago, B. X., Strauss, M. A., Lahav, O., Davis, M.,
Dressler, A., \& Huchra, J. P. 1995, ApJ, 446, 457
\bibitem []{} Strauss, M. A., \& Willick, J. A. 1995, Physics Reports,
261, 271
\bibitem []{} Tonry, J. L., Blakeslee, J. P., Ajhar, E. A., \&
Dressler, A. 1999, astro-ph/9907062
\bibitem []{} Tonry, J. L., \& Davis, M. 1981, ApJ, 246, 666
\bibitem []{} van Albada, T. S., Bertin, G., \& Stiavelli, M. 1993, MNRAS,
265, 627
\bibitem []{} Vorontsov-Velyaminov, B. A., Arhipova, V. P., \&
Krasnogorskaja, A. A. 1962-1974, Morphological Catalog of Galaxies
Vols. 1-5 (Moscow: Moscow State Univ.)(MGC) 
\bibitem []{} Wegner, G., Colless, M., Baggley, G., Davies, R. L.,
Bertschinger, E., Burstein, D., McMahan, R. K., \& Saglia, R. P. 1996,
ApJS, 106, 1
\bibitem []{} Wegner, G., Colless, M., Saglia, R. P.,  McMahan, R. K.,
Davies, R. L., Burstein, D. \&  Baggley, G., 1999, MNRAS, 305, 259
\bibitem []{} Wegner, G., Willmer, C. N. A., Bernardi, M.  \etal, 2000, in preparation
\bibitem []{} Whitmore, B. C., McElroy, D. B., \& Tonry, J. L. 1985, ApJS,
59, 1
\bibitem []{} Willick, J. A. 1990, ApJ, 351, L5
\bibitem []{} Willick, J. A. 1991, PhD Thesis, University of California at
Berkeley
\bibitem []{} Willick, J. A. 1999, ApJ, 522, 647
\bibitem []{} Willick, J. A., Courteau, S., Faber, S. M., Burstein, D.,
Dekel, A., \& Strauss, M. A. 1997, ApJS, 109, 333
\bibitem []{} Willick, J. A., \& Strauss, M. A. 1998, ApJ, 507, 64
\bibitem []{} Zaroubi, S., Zehavi, I., Dekel, A., Hoffman, Y. \&
Kolatt, T. 1997, ApJ, 486, 21
\bibitem []{} Zwicky, F. \etal 1961-1968, Catalog of Galaxies and
Clusters of Galaxies, Vols. 1-6 (Pasadena: California Inst. of Technology)(CGCG) 




\end{thebibliography}
\end{document}